\documentclass[twocolumn,floatfix,prx,aps]{revtex4-1}
\usepackage{graphicx,amsmath,amssymb,color,ulem}
\usepackage{nicefrac}
\usepackage{multirow, makecell, ragged2e}
\usepackage[titletoc,title]{appendix}
\usepackage{bm}

\newcommand{\be}{\begin{equation}}
\newcommand{\ee}{\end{equation}}

\newcommand{\ba}{\begin{eqnarray}}
\newcommand{\ea}{\end{eqnarray}}

\renewcommand{\vec}[1]{\bm{#1}}

\def\beq{\begin{equation}}
\def\eeq{\end{equation}}

\newcommand{\addJKJ}[1]{{\bf \color{magenta} #1}}
\newcommand{\bi}{\bm{i}}
\newcommand{\bj}{\bm{j}}
\newcommand{\bdelta}{\bm{\delta}}

\begin{document}

\title{Topological superconductivity in Landau levels}
\author{Gun Sang Jeon$^{1,2}$, J. K. Jain$^1$, C.-X. Liu$^1$}
\affiliation{$^1$Department of Physics, 104 Davey Lab, The Pennsylvania State University, University Park, Pennsylvania 16802}
\affiliation{$^2$Department of Physics, Ewha Womans University, Seoul 03760, Korea}

\date{\today}

\begin{abstract}
The intense search for topological superconductivity is inspired by the prospect that it hosts Majorana quasiparticles. We explore in this work the optimal design for producing topological superconductivity by combining a quantum Hall state with an ordinary superconductor.
To this end, we consider a microscopic model for
a topologically trivial two-dimensional p-wave superconductor exposed to a magnetic field, and find that the interplay of superconductivity and Landau level physics yields a rich phase diagram of states as a function of $\mu/t$ and $\Delta/t$, where $\mu$, $t$ and $\Delta$ are the chemical potential, hopping strength, and the amplitude of the superconducting gap. In addition to quantum Hall states and topologically trivial p-wave superconductor, the phase diagram also accommodates regions of topological superconductivity. Most importantly, we find that
application of a non-uniform, periodic magnetic field produced by a square or a hexagonal lattice of $h/e$ fluxoids
greatly facilitates regions of topological superconductivity in the limit of $\Delta/t\rightarrow 0$. In contrast,
a uniform magnetic field, a hexagonal Abrikosov lattice of $h/2e$ fluxoids, or a one dimensional lattice of stripes produces topological superconductivity only for sufficiently large $\Delta/t$.
\end{abstract}

\pacs{xxx}

\maketitle

{\it Introduction. -}
Quasi-particle excitations in certain two dimensional (2D) condensed matter systems can possess exotic anyonic statistics that are fundamentally different from the familiar bosonic and fermionic statistics \cite{stern2008anyons,Nayak08}. One example of such quasiparticle excitations is the Majorana zero mode (MZM) \cite{alicea_rpp_2012,beenakker2013search,kitaev2001unpaired}, which obeys non-Abelian statistics and has motivated ideas on topological quantum computation \cite{Kitaev03,Nayak08,sarma2015majorana,alicea2015designer}. MZM can appear as a gapless quasiparticle excitation at the boundary or inside the vortex core of a topological superconductor (TSC). Intensive experimental effort has focused on the realization of TSC and the measurement of MZMs and their non-Abelian statistics in a variety of systems, including 5/2 fractional quantum Hall (QH) state \cite{Read00} (which can be viewed as a TSC of composite fermions), p-wave Sr$_2$RuO$_4$ superconductors (SCs) \cite{rice1995sr2ruo4}, semiconductor nanowires in proximity to SCs under magnetic fields \cite{mourik2012signatures,das2012zero,rokhinson2012fractional,lutchyn2010majorana,sau2010generic,alicea2010majorana,alicea2011non}, magnetic ion chains on top of SC substrates \cite{nadj2014observation}, the surface of (Bi,Sb)$_2$Te$_3$ in proximity to SCs \cite{fu2008superconducting,xu2015experimental,wang2012coexistence}, the surface of Fe(Se,Te) SCs \cite{yin2015observation,wu2016topological,zhang2018observation,wang2015topological} and the heterostructure with a quantum anomalous Hall (QAH) insulator and a SC \cite{qi2010chiral,chung2011conductance,he2017chiral}. In particular, in the last case, recent theory predicted $\frac{e^2}{2h}$ conductance in a junction structure as the transport signature of TSC phase; recent experiments report such a plateau \cite{he2017chiral}, although the interpretation is under debate \cite{lian2018quantum,ji20181,huang2018disorder}. Proposals have been made for demonstrating Majorana braiding and non-Abelian statistics through coherent transport measurement (interferometry) in the QAH-SC hybrid systems \cite{lian2017non}.

Here we consider the possibility of inducing TSC phase in a QH state by coupling it to a topologically trivial SC.
This structure has several differences from the QAH-SC hybrid, with potential advantages as well as challenges.
To begin with, the QH state can readily produce Hall conductance
$\mathcal{C}e^2/h$  with any integer value of the so-called Chern number $\mathcal{C}$, whereas only QAH states  with $\mathcal{C}=\pm 1$ have been observed in experiments (although high-Chern-number QAH states are in principle possible \cite{wang2013quantum,fang2014large}). The possibility of arbitrary $\mathcal{C}$ is thus expected to lead to a richer phase diagram in the QH-SC hybrid. The TSC phase with multiple Majorana edge modes is also of interest because a recent theory \cite{wang2018multiple} has suggested that unique transport signature can arise in a junction with multiple Majorana modes which allows for distinguishing Majorana transport from a trivial interpretation.
Furthermore, the QAH systems are typically highly disordered with low mobilities due to magnetic doping \cite{chang2013experimental}. In contrast, very high mobilities can be achieved for QH samples. The resulting long coherence lengths can be a crucial factor for the success of interference measurements that can possibly verify braiding of emergent anyons. Of course, for coupling between QH state and SC, it would be necessary for superconductivity to contain a non-zero p-wave triplet component, and also to survive to magnetic fields that are sufficiently high to bring the system into QH regime; progress in this direction has been made, as discussed below. Perhaps the most serious conceptual impediment in coupling QH and SC states is that the QH states are gapped states and thus robust to small perturbations, including coupling to weak superconductivity. A crucial step of this work is to show that one may circumvent this problem by considering spatially non-uniform magnetic fields to produce ``dispersive" LLs, which support gapless states for appropriate chemical potentials. We further find that the feasibility of TSC depends also on the magnetic flux lattice structure, and identify which geometries are most hospitable to TSC in the weak coupling limit.

The topological character of the states of interest to us will be quantified by the Chern number. The QH state has Chern number $\mathcal{C}=$integer, which corresponds to the number of LLs below the chemical potential. We will use the Bogoliubov-de Gennes (BdG) framework to enable a treatment of superconductivity.
In the BdG formulation, we can define another Chern number $\mathcal{N}$, defined below, which is called SC Chern number. In the zero SC gap limit, a QH state has $\mathcal{N}=2\mathcal{C}$ due to the redundancy of the BdG Hamiltonian. Therefore, any state with even $\mathcal{N}$ is equivalent to a QH state and not of interest to us.
We define below the states with odd integer values of $\mathcal{N}$ as the TSC states.
Such a state should possess an odd number of chiral Majorana modes at the boundary and odd number of MZMs trapped at the core of a vortex. Non-Abelian statistics for the vortex is only possible for a state with odd $\mathcal{N}$.

In what follows, we explore a microscopic model of 2D spinless electron system with nearest-neighbor SC pairing subjected to a magnetic field, and obtain the phase diagram of various states as a function of the chemical potential ($\mu$), the hopping strength ($t$) and magnitude of the superconducting gap ($\Delta$).
For a uniform magnetic field, we find, as expected, that a TSC phase requires the strength of the superconducting coupling to be enhanced beyond a critical value to overcome the QH gap. We therefore consider non-uniform magnetic fields, which still produce LLs, but the LLs are dispersive (rather than flat). The most natural model, that of an Abrikosov lattice of $h/2e$ fluxoids, fails to produce TSC for small $\Delta/t$; see Supplementary Material (SM)\cite{Jeon-SM}.
In contrast, a square lattice of $h/e$ fluxoids produces an enormously rich phase diagram with many regions of TSC with various odd $\mathcal{N}$. Most importantly, even a weak p-wave superconductor ($\Delta/t\rightarrow 0$) can turn into a TSC for appropriate chemical potentials. We have also considered a geometry where the magnetic field forms stripes; this geometry fails to produce TSC for small $\Delta/t$.
Our calculations thus provide insight into how best to integrate SC gap with LLs to realize TSC. We also discuss possible experimental manifestations.

\begin{figure}
	\includegraphics[width=0.4\textwidth]{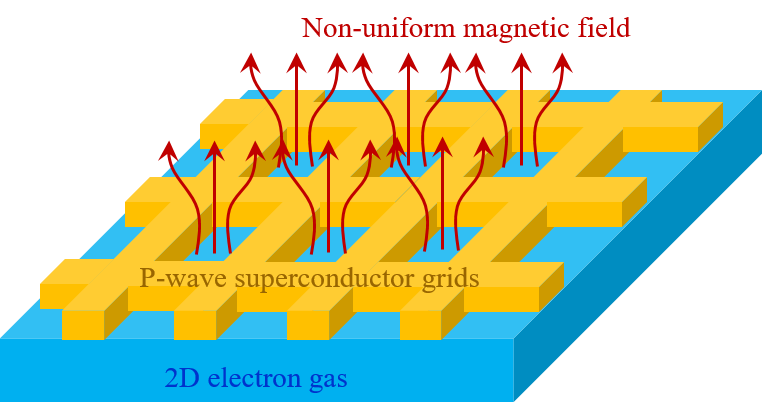}
	\centerline{(a)}
	\includegraphics[width=0.17\textwidth]{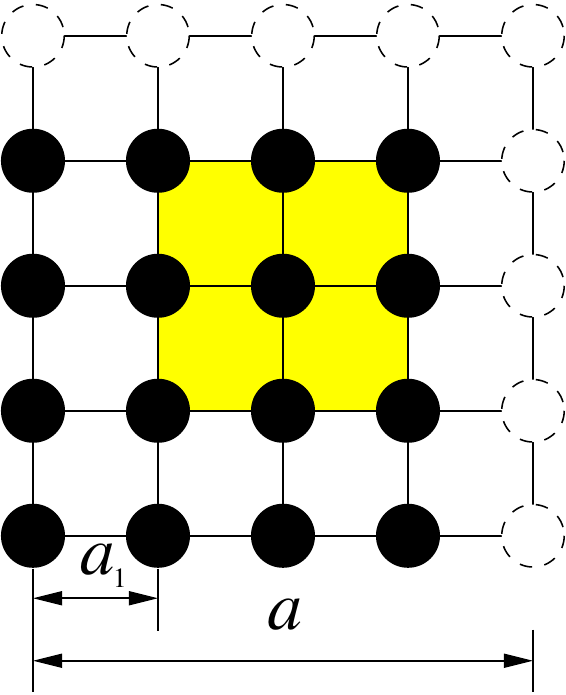}
\hspace*{5ex}
	\includegraphics[width=0.17\textwidth]{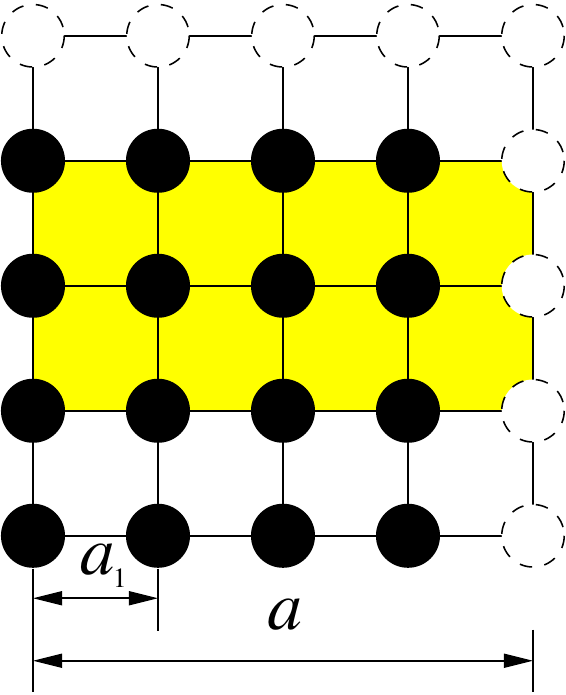}
	\centerline{(b) \hspace*{25ex}  (c)}
	\caption{
		\label{fig:schematic}
		(a) Experimental setup for a hybrid system with two-dimensional electron gas with $p$-wave superconducting grid on top under an external magnetic field.
		(b-c) Schematic of a unit cell of $N\times N$ lattice points containing one flux quantum (b) of a square shape; (c) in stripe geometry.
		In each unit cell a flux quantum $\phi_0$ penetrates through the yellow shaded zone of (b) $M\times M$ (c) $N\times M$ lattice points. The figures represent $N=4$ and $M=2$.
		The size of unit cell is $a$ and spacing between lattice points is $a_1=a/N$.
	}
\end{figure}

\begin{figure}
	\includegraphics[width=0.45\textwidth]{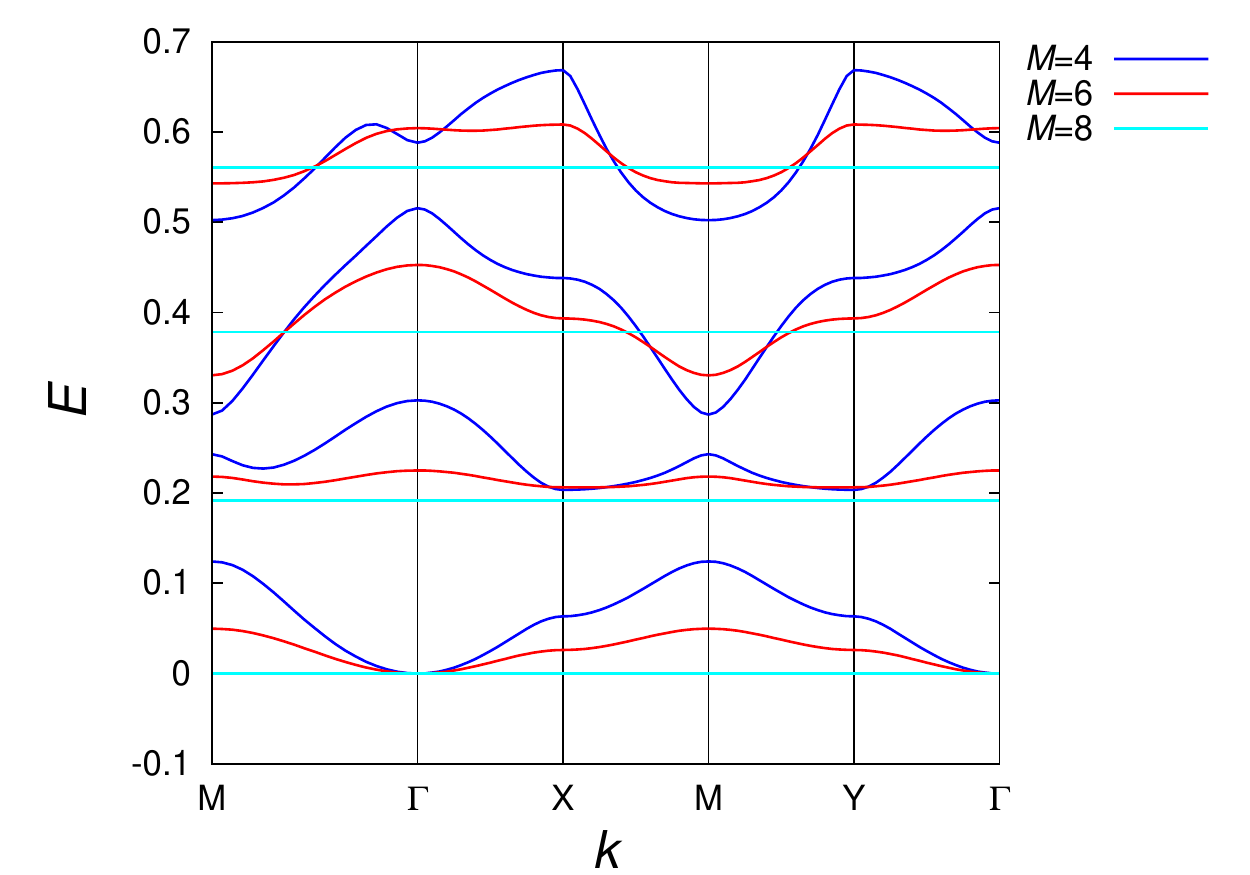}
	\centerline{(a)}
	\includegraphics[width=0.45\textwidth]{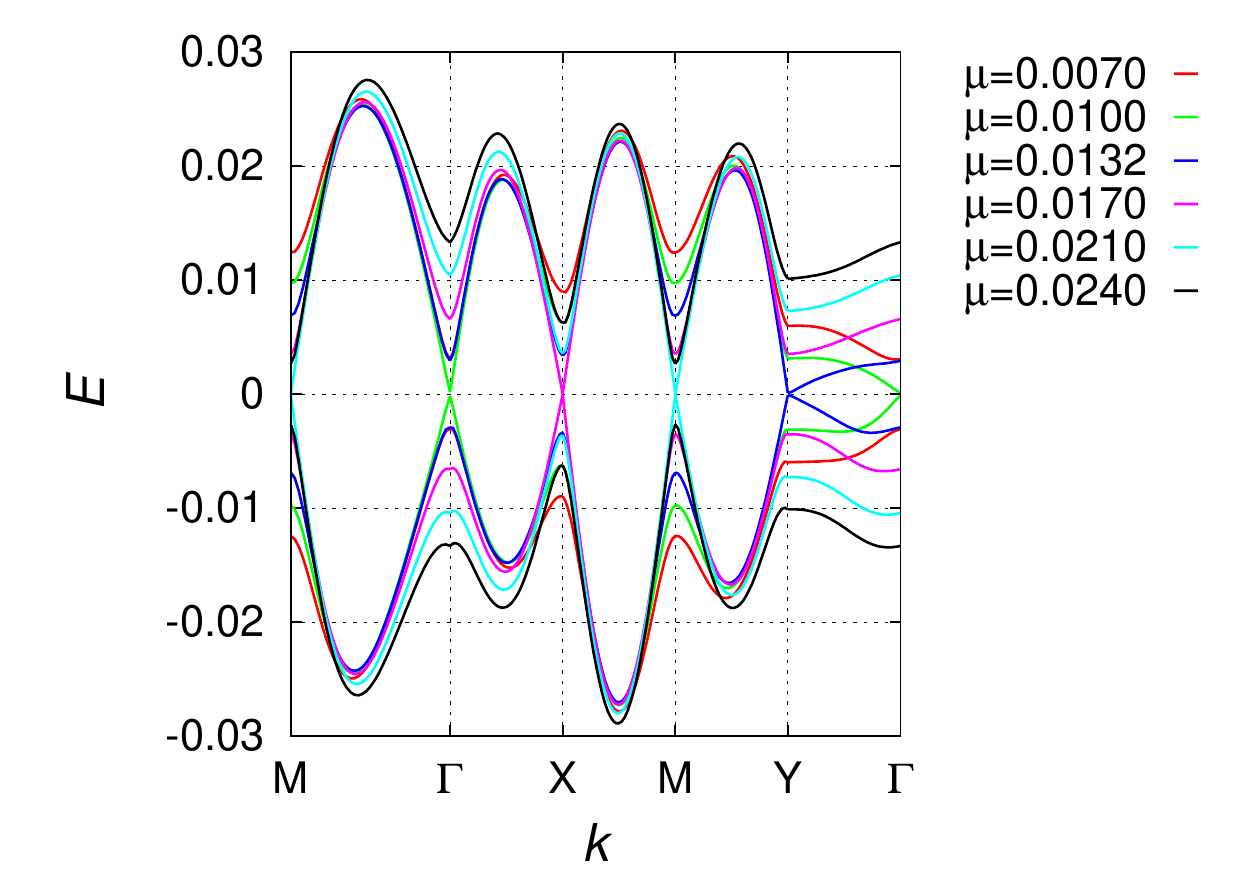}
	\centerline{(b)}
	\caption{
		(a) Energy dispersion in the presence of uniform magnetic field ($M=8$) for $N=8$ and $\Delta=0$.
		(b) Energy dispersion near the Fermi energy for various values of chemical potential $\mu$ in the presence of uniform magnetic field ($M=8$) with $N=8$ and $\Delta=0.1$.
	}
\end{figure}
{\it Model Hamiltonian -} We have in mind a general flux lattice, which can possibly be realized through a hybrid of 2D electron gas with p-wave SC grid on top exposed to an external magnetic field [Fig.~\ref{fig:schematic}(a)]. (The p-wave SC grid serves a dual purpose: it induces superconductivity and also produces a magnetic flux lattice.)
We model such a hybrid system through spinless fermions on a square lattice in the presence of nearest-neighbor pairing and a perpendicular magnetic field. (The lattice represents the continuum in the limit of vanishing lattice spacing. We have confirmed, as elaborated in the SM~\cite{Jeon-SM}, that the results presented below provide a good qualitative and semi-quantitative approximation of the continuum limit.)
One unit cell consists of $N\times N$ lattice sites with lattice spacing $a_1$, as shown in Fig.~\ref{fig:schematic}(b) and (c).
We next show results for a uniform magnetic field as well as for several kinds of flux lattices.
In the first configuration, depicted in Fig.~\ref{fig:schematic}(b),
single flux quanta ($h/e$) of a square shape are located in an $M\times M$ square at the center of each unit cell.
Figure~\ref{fig:schematic}(c) shows single flux quanta in an $N\times M$ rectangle in each unit cell,
producing periodic stripes of magnetic field.
In both configurations $M=N$ gives uniformly distributed magnetic field. (It is noted that the $h/e$ flux quantum must have a finite extent to have an effect; a point flux quantum may be gauged away.)
The lattice periodicity is $a=N a_1$ and the lattice vector
is decomposed as
$\vec{r}_{\bm{i}=(i_x,i_y)} = \vec{R}_{\bm{m}=(m_x,m_y)} + \vec{\tilde{r}}_{\bm{l}=(l_x,l_y)}$
for $ (i_x,i_y) = (N m_x + l_x,N m_y + l_y)$ with
$0\le l_{x/y} < N$, $0\le m_{x/y} < L $ , and $L a$ being the linear size of the system.
Here,
$\vec{R}_{\bm{m}}$ represents the position of the reference point in the $\bm{m}$th unit cell and $\vec{\tilde{r}}_{\bm{l}}$ is the internal position relative to $\vec{R}_{\bm{m}}$.

The full Hamiltonian with nearest neighbor pairing on the lattice is given by
\begin{align}
	\mathcal{H} &= -t \sum_{\bj,\bdelta} ( e^{i A_{\bj+\bdelta,\bj}} c_{\bj+\bdelta}^\dagger c_{\bj} + e^{i A_{\bj,\bj+\bdelta}} c_{\bj}^\dagger c_{\bj+\bdelta} ) \nonumber\\
											  &-\mu \sum_{\bi} c_{\bi}^\dagger c_{\bi} - \Delta \sum_{\bj,\bdelta} (c_{\bj+\bdelta} c_{\bj} + c_{\bj}^\dagger c_{\bj+\bdelta}^\dagger)
\end{align}
where
$c_{\bi}$ ($c_{\bi}^\dagger$) annihilates (creates) a spinless electron at $\vec{r}_{\bi}$;
$\bdelta= (1,0),(0,1)$; $A_{\bj,\bj'}=-A_{\bj',\bj}$; and
\begin{align}
\sum_P A_{\bj, \bj'} =
\left\{
\begin{array}{cc}
2\pi f & \hbox{inside a flux}
\\
0 & \hbox{otherwise}
\end{array}
\right.
\end{align}
with $P$ denoting a directional plaquette sum and $f \equiv 1/N_P$.
Here $N_P$ is the number of plaquettes though which fluxes penetrate in one unit cell. ($N_P=M^2$ for square-shape flux and $N_P=N M$ for stripe flux.)
With
$c_{\bi} = \frac{1}{L} \sum_{\vec{k}} e^{i \vec{k} \cdot \vec{R_\vec{m}}} c_{\vec{k},\vec{\tilde{r}}_{\vec{l}}}$,
where $|k_x|\le \pi/a, |k_y|\le \pi/a$,
we can transform the Hamiltonian into the BdG form
on the basis
$\vec{C}_{\vec{k}}^\dagger \equiv
(
c_{\vec{k},\vec{\tilde{r}}_{(0,0)}}^\dagger,
 \ldots,
 c_{\vec{k},\vec{\tilde{r}}_{(N{-}1,N{-}1)}}^\dagger,
 c_{-\vec{k},\vec{\tilde{r}}_{(0,0)}},
 \ldots,
 c_{-\vec{k},\vec{\tilde{r}}_{(N{-}1,N{-}1)}}
)$.
The detailed matrix elements are given in the SM.
The numerical diagonalization of the BdG Hamiltonian gives $2N^2$ bands of eigen-energies $E_{n,\vec{k}}$ with the eigenvectors $|n,\vec{k}\rangle$ with $n=1,2,\ldots,2N^2$.
The SC Chern number $\mathcal{N}$ can be computed by \cite{Thouless82,qi2010chiral}
\begin{align} \nonumber
	\mathcal{N} =
	&
	\frac{1}{\pi}
	\int d^2\vec{k}
	\sum_{n=1}^{N^2}
	\sum_{m(\ne n)}^{2N^2}
	\\
	& \quad
	\frac{ {\rm Im}
		\langle m,\vec{k} | (\partial_{\vec{k}_x} \mathcal{H}_{\vec{k}}) | n,\vec{k} \rangle^*
		\langle m,\vec{k}| (\partial_{\vec{k}_y} \mathcal{H}_{\vec{k}}) |n,\vec{k} \rangle
		 }{
		 (E_{n,\vec{k}} - E_{m,\vec{k}})^2
	 	},
\end{align}
and identifies each topological phase in the phase diagram.
The Chern number $\mathcal{N}$ changes at the boundaries of two distinct phases where the energy gap closes at some momentum.
Henceforth, the energy is expressed in units of $t$. (Note that we have defined the SC Chern number with respect to half of the bands, rather than with respect to occupied states below the chemical potential. In cases where the chemical potential lies inside a gap, the two definitions coincide. We also find some semi-metallic regions\cite{Jeon-SM} where this is not the case.)

\begin{figure}
	\includegraphics[width=0.45\textwidth]{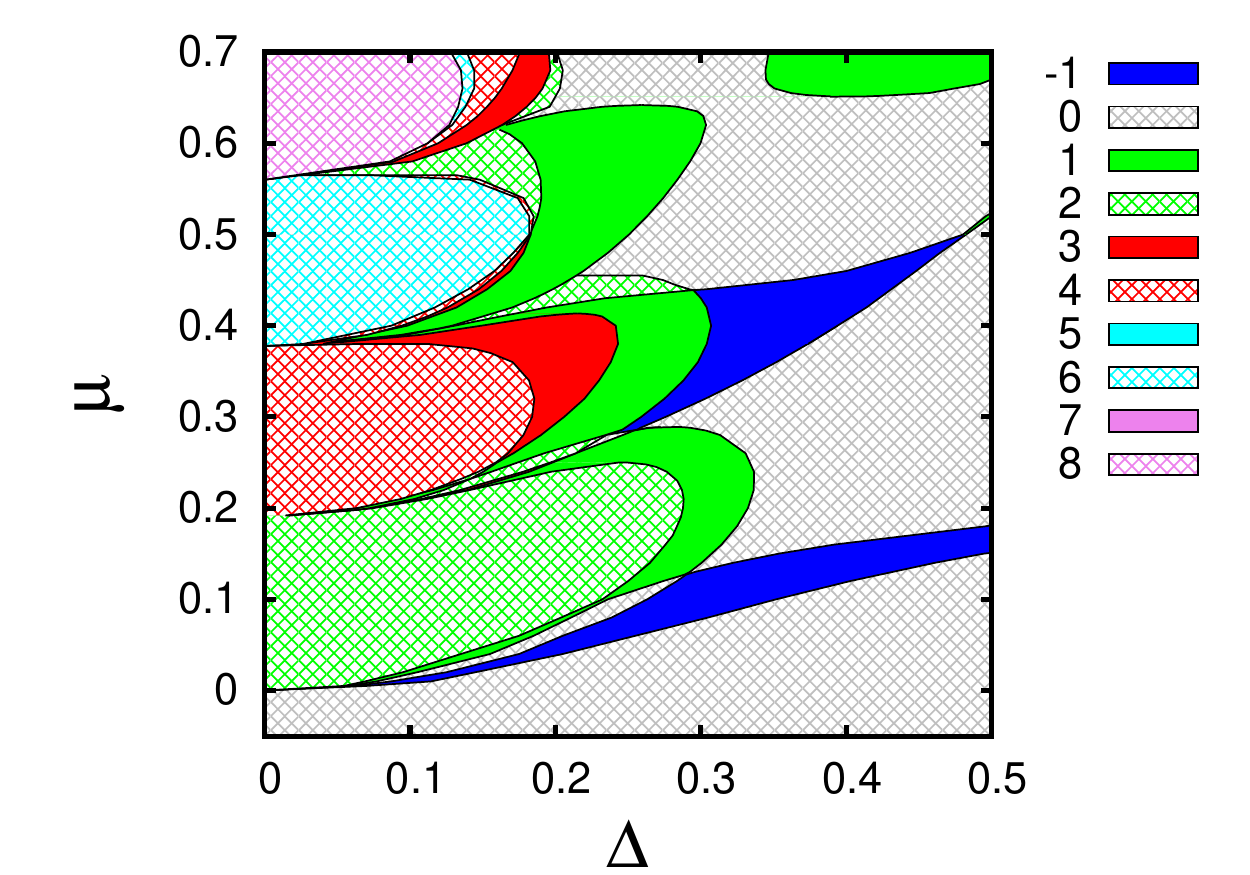}
	\centerline{(a)}
	\includegraphics[width=0.45\textwidth]{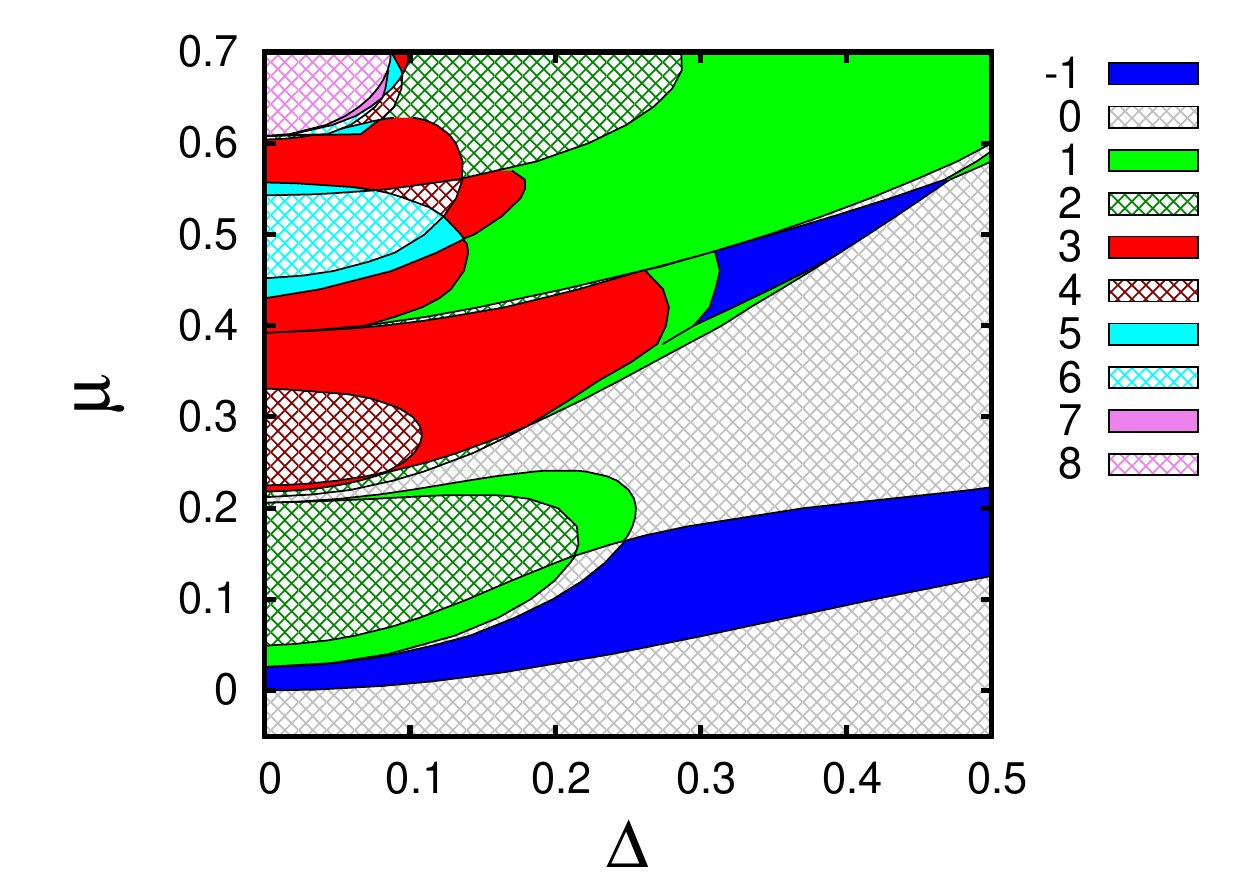}
	\centerline{(b)}	
	\includegraphics[width=0.45\textwidth]{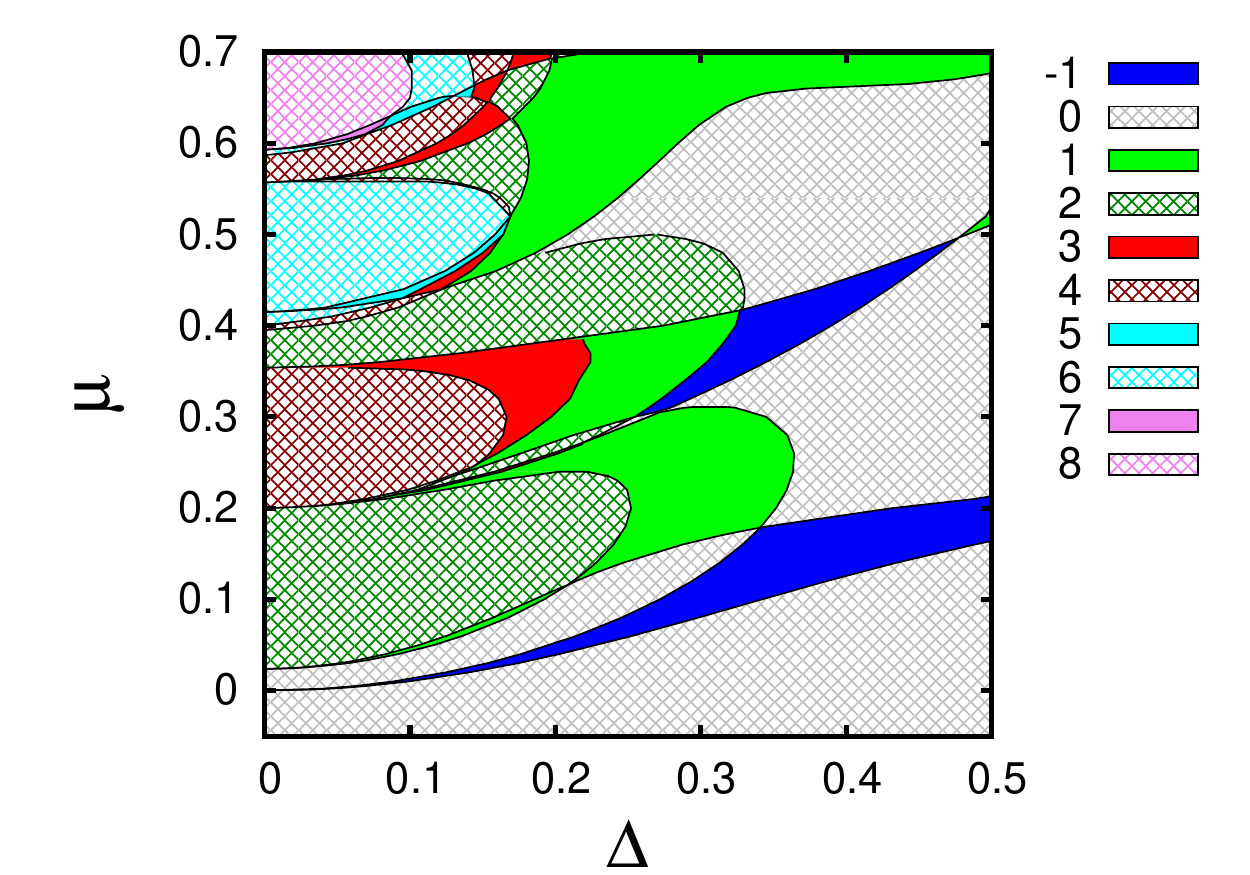}
	\centerline{(c)}
	\caption{
			  Phase diagram of states for various geometries. Each phase is characterized by its Chern number ${\cal N}$, with different Chern numbers shown in different colors. The gray hashed region is an insulator; the colored hashed regions represent even integer ${\cal N}$, which correspond to QH effect; and the solid colors depict the topological superconductor phase with odd Chern numbers. The different panels show:
			  (a) a uniform magnetic field ($N=8,M=8$),
			  (b) a non-uniform periodic magnetic field of a square shape ($N=8,M=6$), and (c) a non-uniform magnetic field in stripe geometry ($N=8,M=6$).
	}
\end{figure}

{\it TSC phase and Phase Diagram:}
The energy dispersion of our model Hamiltonian without superconductivity ($\Delta=0$) is shown in Fig. 1a for different magnetic flux sizes $M$. A uniformly distributed magnetic field ($M=N$) yields flat Landau levels (cyan lines in Fig. 1a).
For non-uniform magnetic fields, Landau levels turn into dispersive "Landau" bands (blue and red lines in Fig. 1a). The Chern number $\mathcal{C}$ carried by each band is unchanged since there is no level crossing between different bands. Due to the dispersive bands, the Fermi surface can appear and the system becomes metallic for certain ranges of $\mu$, in contrast to the insulating phase of filled Landau levels obtained for a uniform magnetic field. A SC gap can be opened by turning on $\Delta$. By varying the Fermi energy, multiple Dirac type of transitions are found, indicating the existence of topological phase transitions.
In Fig.2, we plot variation of energy dispersions for several values of chemical potential $\mu$ with $\Delta=0.1$ in the presence of uniform magnetic fields ($N=8,M=8$).
At $\mu=0$ the system lies in a topologically trivial phase.
As $\mu$ increases, we observe successive gap closings at symmetry points $\Gamma$, $\hbox{Y}$, $\hbox{X}$, and
$\hbox{M}$.
Each gap closing accompanies a unit change in $\mathcal{N}$, and accordingly the system is expected to exhibit two TSC phases, one between the gap closings at $\Gamma$ and Y and the other between those at X and M. The numerical computation of $\mathcal{N}$ reveals that the former corresponds to $\mathcal{N}=-1$ and the latter to $\mathcal{N}=1$.
After the last gap closing at M, the SC Chern number is 2, corresponding to $\mathcal{C}=1$ QH state.

We numerically evaluate the SC Chern number together with the gap closing momenta in a wide region of $(\Delta,\mu)$.
The resulting phase diagram of SC Chern number as a function of the Fermi energy $\mu$ and SC gap $\Delta$ is shown in Fig. 3 for $N=M$ (uniform $B$) as well as $N\neq M$ (non-uniform $B$). The regions of even $\mathcal{N}$, indicated by hashes, are adiabatically connected, and thus equivalent, to conventional QH states with filling factor $\mathcal{C}=\frac{\mathcal{N}}{2}$. The regions with solid colors depict SC states with odd integer values of $\mathcal{N}$, i.e. the TSC phase. We also find narrow regions of semi-metallic phase near the phase boundaries where $\mathcal{N}$ changes by 2. These regions are discussed in the SM\cite{Jeon-SM}, but suppressed in Fig.~3 to avoid clutter.

It is evident that the phase diagram is very sensitive to how the magnetic field penetrates the superconductor. For a uniform magnetic field (Fig. 3a),
only even values of $\mathcal{N}$ appear in the limit of weak SC gap $\Delta/t\rightarrow 0$, although TSC phases with an odd $\mathcal{N}$ can occur when $\Delta$ is sufficiently large. The same is true of the stripe geometry (Fig.~3c). As discussed in the SM~\cite{Jeon-SM}, a hexagonal lattice of $h/2e$ fluxoids is also not effective in producing TSC.
In striking contrast, for the non-uniform magnetic field produced by a square lattice of $h/e$ fluxoids, TSC phases can emerge even in the limit $\Delta/t\rightarrow 0$, as shown by the blue, red and dark green areas in Fig. 3b. This property is also shared by a hexagonal lattice of $h/e$ fluxoids~\cite{Jeon-SM}. We thus conclude that a square or a hexagonal lattice of $h/e$ fluxoids is the best geometry for generating TSC.

{\it Discussion and Conclusion:}
Mong {\it et al.} \cite{Mong14} considered stripes of QH states with their oppositely moving chiral edge states coupled by SC coupling or tunneling, and demonstrated emergence of TSC for certain parameters. While our work is in a topological sense similar, our model does not have edge states, and both QH effect and superconductivity coexist throughout the entire sample. Our study also allows high filling factors, thereby producing a rich phase diagram. Our model suggests that TSC phases can generally exist in QH-SC hybrid, without requiring fine-tuning of the parameters, provided the magnetic field has periodic spatial variation.

Before ending the article, it is appropriate to discuss potential experimental realizations of the TSC phase and its transport signatures.
An experimental challenge towards investigating this physics was that the conditions for producing QH effect and superconductivity appear incompatible: high magnetic fields required for the QH effect are inimical to  superconductivity. Important experimental progress has recently been made in this direction. Supercurrent and Josephson coupling in QH regime at graphene-superconductor interface have been demonstrated at relatively low magnetic field ($\sim$ 2T)\cite{Rickhaus12,Amet16,Shalom16}. In another work, superconducting niobium nitride (NbN) electrode with very high critical magnetic field has been used to induce superconducting correlations in the QH edge states of graphene to see evidence of crossed Andreev reflection on a QH plateau\cite{Lee17}.
In yet another work, inter-Landau level Andreev reflection has been observed in graphene coupled to an NbSe$_2$ superconductor~\cite{Sahu18}. While the superconductor contains spin singlet Cooper pairs, it is in principle possible to induce spin triplet superconductivity by exploiting spin-orbit coupling or inhomogeneous magnetization at the interface. Such phenomena have been studied in a variety of hybrid systems with SC and magnetic or strong spin-orbit coupled materials \cite{buzdin2005proximity,bergeret2005odd,eschrig2011spin,eschrig2015spin}. The feasibility to implement such mechanism in QH-SC hybrid will be addressed in a future work. It should be possible
to construct a variety of QH-TSC planar junctions \cite{lian2017non,chung2011conductance,lian2016edge,wang2018multiple} where the chemical potential may be controlled through local gates. One advantage is the possibility of achieving the TSC phase with $\mathcal{N}>1$ in junction structures, which has theoretically been proposed to give rise to unique transport signature \cite{wang2018multiple} that can unambiguously establish chiral Majorana transport.

In summary, we have identified optimal conditions for producing a topological superconductor by studying coupling between QH effect and superconductivity in a microscopic model. We find that a non-uniform magnetic field produced by a square or a hexagonal lattice of $h/e$ fluxoids is likely the best geometry for this purpose.

{\it Acknowledgement -}
We acknowledge the support from Office of Naval Research under Grant No. N00014-15-1-2675 (CXL), by the U. S. Department of Energy under Grant no. DE-SC0005042 (JKJ), and by the National Research Foundation of Korea under Grant No. NRF-2018R1D1A1B07048749 (GSJ).

\newcommand{\btau}{\bm{\tau}}
\newcommand{\bl}{\bm{l}}
\newcommand{\bk}{\bm{k}}

\begin{widetext}
	\begin{center}

\textbf{\large Supplementary Material for: ``Topological superconductivity in Landau levels"}
	\end{center}

\setcounter{figure}{0}
\setcounter{equation}{0}
\renewcommand\thefigure{S\arabic{figure}}
\renewcommand\theequation{S\arabic{equation}}
\renewcommand\thesection{\arabic{section}}

\section{Matrix Elements of BdG Hamiltonian}

We give here the detailed form of the BdG Hamiltonian for a 2D electron gas in a magnetic field.
With the Fourier transform
$c_{\bi} = \frac{1}{L} \sum_{\vec{k}} e^{i \vec{k} \cdot \vec{R_i}} c_{\vec{k},\vec{\tilde{r}}_{\bi}}$,
where $|k_x|\le \pi/a, |k_y|\le \pi/a$,
we obtain the Hamiltonian in the BdG form
\begin{align}
&\mathcal{H} = \sum_{\vec{k}} \vec{C}_{\vec{k}}^\dagger \vec{H}_{\vec{k}} \vec{C}_{\vec{k}},\nonumber\\
&\vec{H}_{\vec{k}}
=
\left(
		\begin{array}{cc}
			\vec{T}_{\vec{k}} & \vec{\Delta}_{\vec{k}}^\dagger
			\\
			\vec{\Delta}_{\vec{k}} & -\vec{T}_{-\vec{k}}^\dagger
		\end{array}
\right)
\end{align}
where
$\vec{C}_{\vec{k}}^\dagger =
(
 c_{\vec{k},\vec{\tilde{r}}_1}^\dagger,
 \ldots,
 c_{-\vec{k},\vec{\tilde{r}}_1},
 \ldots
)$.
The nonzero diagonal and off-diagonal terms in the above BdG Hamiltonian
are
\begin{align}
(\vec{T}_{\vec{k}})_{\vec{\tilde{r}},\vec{\tilde{r}}'} &=
\left\{
\begin{array} {ll}
	-\mu & \hbox{for } \vec{\tilde{r}'}=\vec{\tilde{r}}
	\\
- t e^{i A_{\vec{\tilde{r}},\vec{\tilde{r}}'}} & \hbox{for } \vec{\tilde{r}'}=\vec{\tilde{r}} + a_1 \hat{x}, \vec{\tilde{r}} + a_1 \hat{y}
	\\
	- t e^{i (A_{\vec{\tilde{r}},\vec{\tilde{r}}'}+k_x a)} & \hbox{for } \vec{\tilde{r}'}=\vec{\tilde{r}}- (N-1)a_1\hat{x}
	\\
	- t e^{i (A_{\vec{\tilde{r}},\vec{\tilde{r}}'}+k_y a)} & \hbox{for } \vec{\tilde{r}'}=\vec{\tilde{r}}- (N-1)a_1\hat{y}
	\\
\end{array}
\right.
\\
({\vec{\Delta}}_{\vec{k}})_{\vec{\tilde{r}},\vec{\tilde{r}}'} &=
\left\{
\begin{array}{ll}
	 - \Delta & \hbox{for } \vec{\tilde{r}'}=\vec{\tilde{r}} + a_1 \hat{x} \hbox{ or }\vec{\tilde{r}} + a_1 \hat{y}
	\\
	 - \Delta e^{i k_x a} & \hbox{for } \vec{\tilde{r}'}=\vec{\tilde{r}}- (N-1)a_1\hat{x}
	\\
	 - \Delta e^{i k_y a} & \hbox{for } \vec{\tilde{r}'}=\vec{\tilde{r}}- (N-1)a_1\hat{y}
\end{array}
\right.
\end{align}
and their transposed elements given by
$(\vec{T}_{\vec{k}})_{\vec{\tilde{r}}',\vec{\tilde{r}}} = (\vec{T}_{\vec{k}})_{\vec{\tilde{r}},\vec{\tilde{r}'}}^*$ and
		  $(\vec{\Delta}_{\vec{k}})_{\vec{\tilde{r}}',\vec{\tilde{r}}} = - (\vec{\Delta}_{\vec{k}})_{\vec{\tilde{r}},\vec{\tilde{r}'}}^*$.

\section{Certain general properties of the phase diagrams}

 \begin{figure}[b]
	\parbox{0.45\textwidth}{
 	\includegraphics[width=0.45\textwidth]{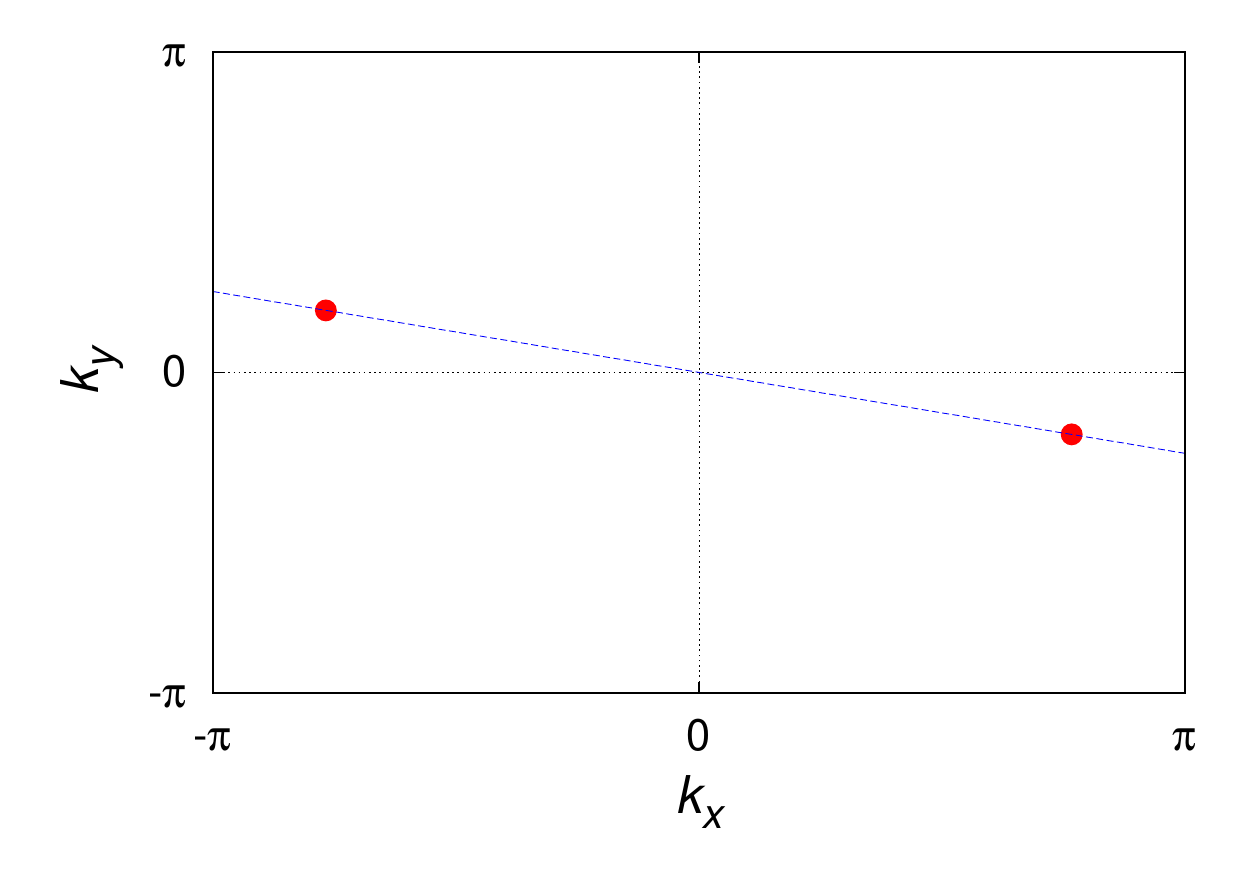}
 	\centerline{(a)}
}
	\parbox{0.45\textwidth}{
	 \includegraphics[width=0.45\textwidth]{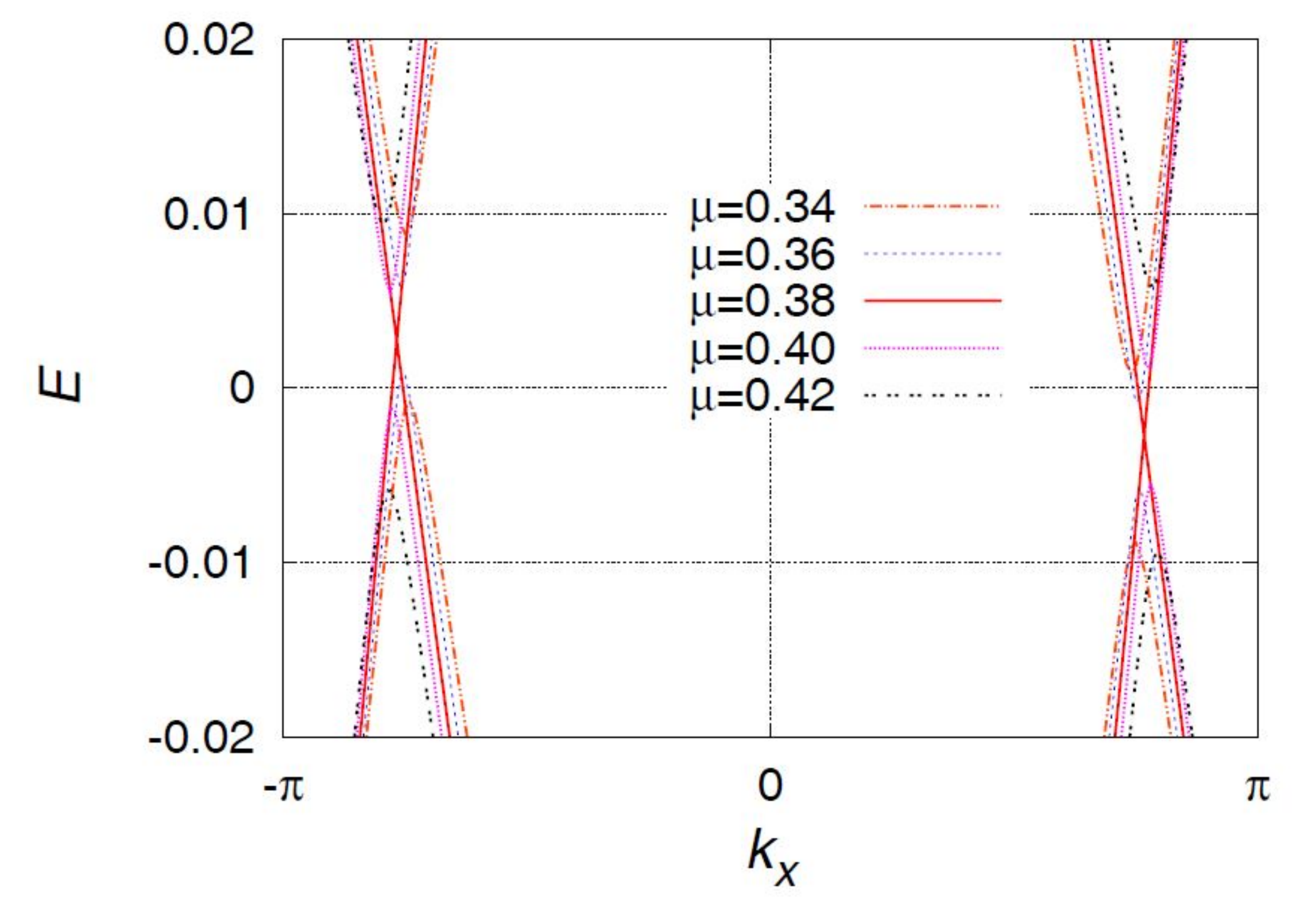}
 	\centerline{(b)}
}
	\caption{ \label{fig:twofold}
		(a) When a gap closes at a momentum $\bm{k}_o$ that does not lie at a symmetry point, the energy dispersion at $-\bm{k}_o$ also undergoes simultaneous gap closing. Consequently, Chern number changes by two on the phase boundary. The two red solid circles indicate the two momenta $\bm{k}_o$ and $-\bm{k}_o$ where the electron and the hole bands touch simultaneously for $\mu=0.38$ and $\Delta=0.243$ with $N=M=8$.
		(b) The energy dispersions along the blue dashed line in (a) for $\Delta=0.243$ and several values of $\mu$  with $N=M=8$. The semi-metal phase with electron and hole pockets occurs at $\mu=0.38$ (which is the phase boundary where $\mathcal{N}$ changes by 2) and $\mu=0.36$.
 	}
 \end{figure}

\begin{figure}
	\parbox{0.45\textwidth}{
	\includegraphics[width=0.45\textwidth]{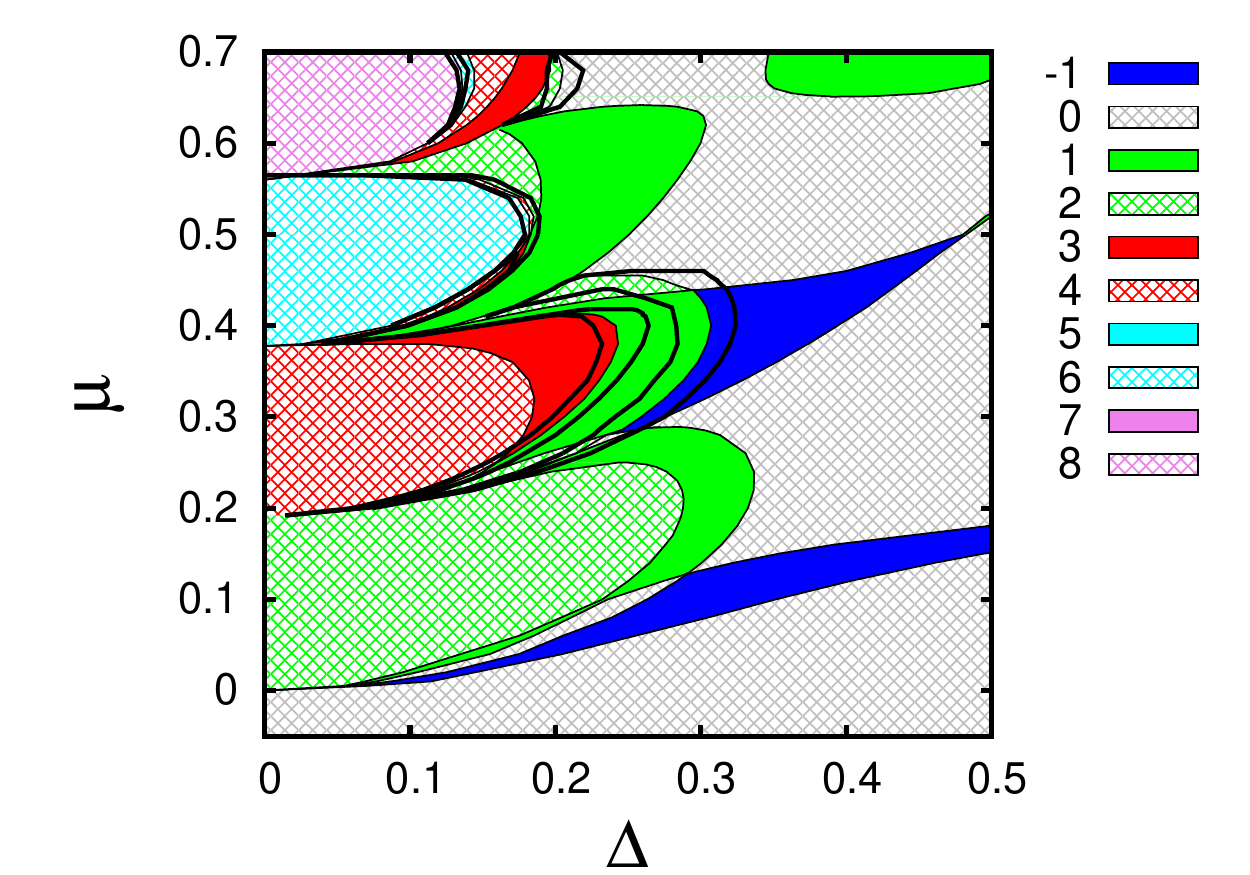}
	\centerline{(a)}
}
	\parbox{0.45\textwidth}{
	\includegraphics[width=0.45\textwidth]{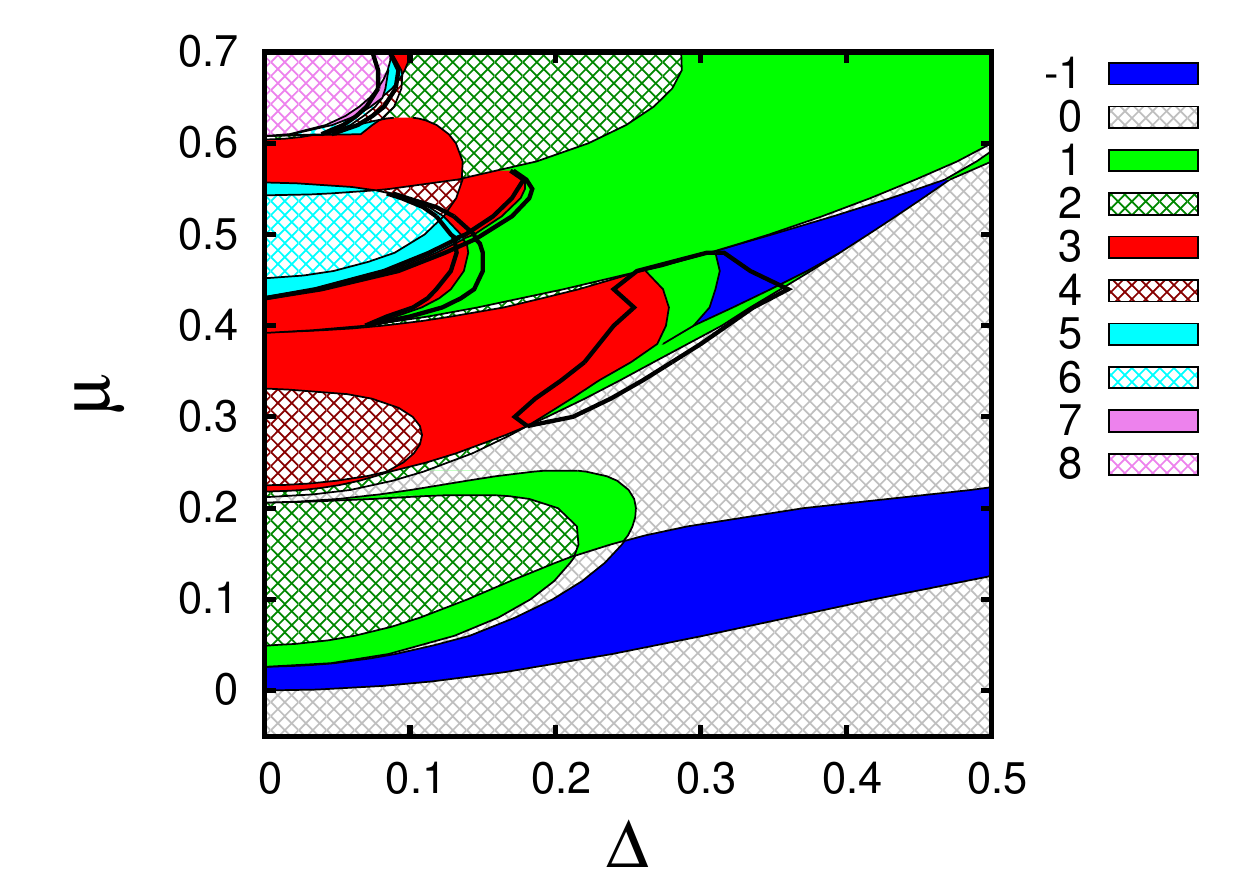}
	\centerline{(b)}
}
	\caption{ \label{fig:PD_sq_metal}
			  Phase diagram under
			  (a) uniform magnetic field ($N=8,M=8$);
			  (b) non-uniform magnetic field ($N=8,M=6$).
			  On the boundary when Chern number changes by two, we observe a semi-metallic phase on the finite region, which is displayed by black thick solid lines.
	}
\end{figure}
		
In our calculated phase diagrams, the SC Chern number $\mathcal{N}$ generally decreases, going through successive transitions, as $\Delta$ is increased. Thus, conventional quantum Hall states with high $\mathcal{N}$, which exist for small $\Delta$, is transformed to a trivial superconductor with $\mathcal{N}=0$, which is expected to occur at extremely large $\Delta$.

Whether $\mathcal{N}$ changes by one or two depends on whether the gap closes at one of the symmetry points or not. $\mathcal{N}$ changes by one on the boundaries where an energy gap closes at one of the symmetry points. Otherwise, $\mathcal{N}$ changes by two because the gap closes simultaneously
at two different momenta $\bm{k}_o$ and $-\bm{k}_o$, where $\bm{k}_o$ lies away from any high symmetry momentum.
Figure~\ref{fig:twofold}(a) shows an example of gap closing at momenta that do not lie at symmetry points for $\mu=0.38$ and $\Delta=0.243$ with $N=M=8$.
On this boundary the system undergoes a transition from $\mathcal{N}=3$ to $\mathcal{N}=1$.

 When the SC Chern number $\mathcal{N}$ changes by one at high symmetry momentum, the gap closing always occurs at the chemical potential. However, when the SC Chern number $\mathcal{N}$ changes by two, the energy gap does not close exactly at the Fermi level, due to the absence of both time reversal and inversion symmetries in our model Hamiltonian. As shown in Fig.~\ref{fig:twofold}(b), the gap closes at an energy slightly higher than the Fermi level ($E>0$) at one momentum $\bm{k}_o$ while slightly below the Fermi energy ($E<0$) at the opposite momentum $-\bm{k}_o$.  This indicates that there exists, in some parameter regions on either side of the phase boundary, a semi-metallic phase with hole and electron pockets at two opposite momenta, although two bands close to the Fermi level do not touch each other. (As noted in the main text, we have defined the SC Chern number with respect to half of the bands in the BdG formalism, rather than with respect to occupied states below the chemical potential. The two definitions being identical when the chemical potential lies inside a gap, but not when the chemical potential crosses a band. In the semi-metallic region, while we find an integer value for $\mathcal{N}$ with our definition, this value does not correspond to a physically measurable transport coefficient.) The semi-metallic regions are often small and were omitted in the phase diagrams shown in the main text to avoid distraction from the physics of TSC. For completeness, in  Figs.~\ref{fig:PD_sq_metal} (a) and (b), the semi-metallic regions are demonstrated by the black thick closed lines.

\section{Continuum Limit}
We discuss the effects of a finite number $N$ of lattice in a unit cell.
The number of lattice points in a unit cell is $N^2$, which produces $N^2$ bands in the finite energy windows $4t$. We fix the magnetic flux in one unit-cell to be one flux quantum while varying $N$.
The correct continuum limit is expected to maintain the constant spacing between Landau levels in the scaled energy of $N^2 E$.
\addJKJ{} Figure~\ref{fig:sE}(a) demonstrates that the scaled energy measured from the minimum of the lowest band almost the same Landau level spacing for different $N$. The first few lowest levels exhibit essentially the same scaled energy even for $N=8$.
Although some quantitative deviation occurs for higher levels, they also saturate as $N$ is increased.

In the presence of a nonuniform magnetic field, the scaled energies display similar saturations with the increase of $N$.
In particular, the system with $N=8$ already captures the essential features of the dispersive bands of the continuum limit.

\begin{figure}
	\parbox{0.45\textwidth}{
	\includegraphics[width=0.45\textwidth]{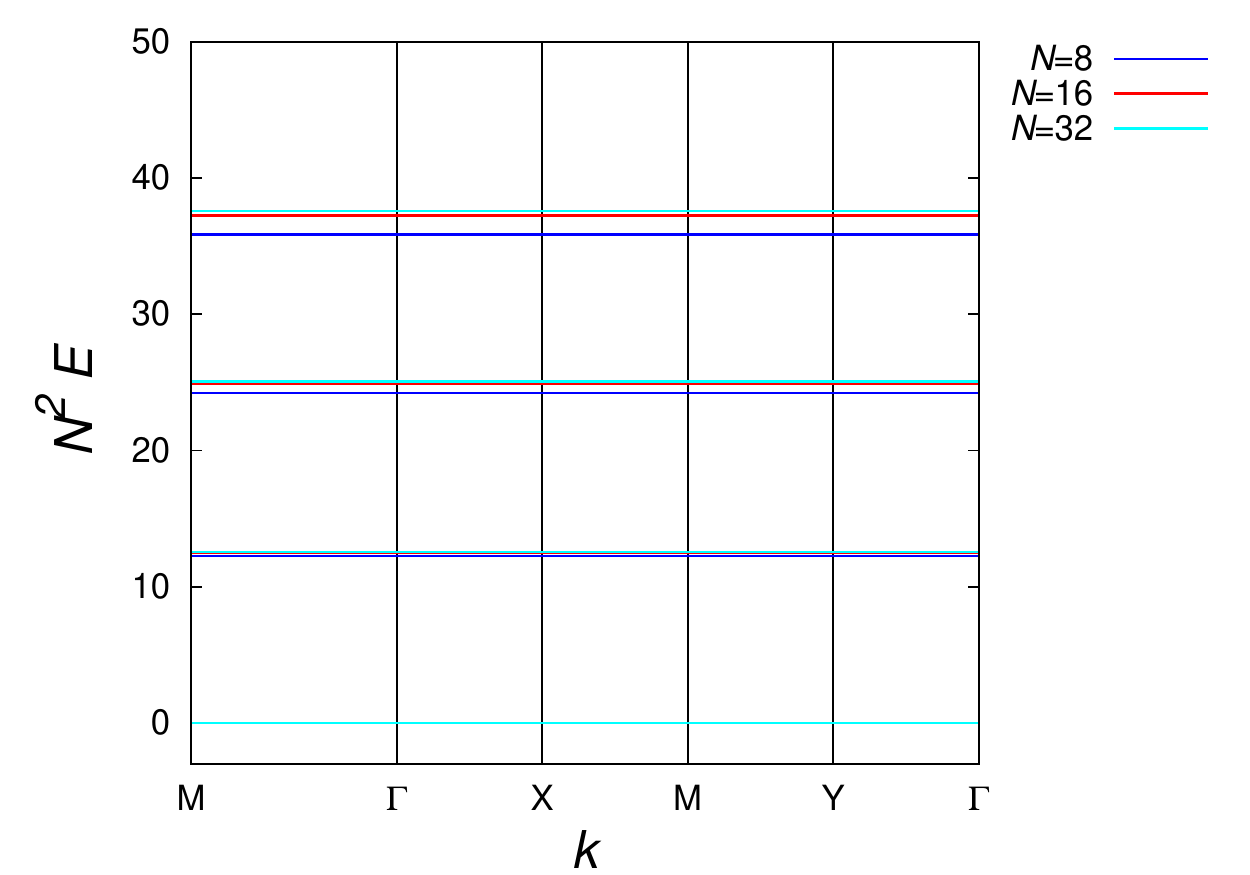}
	\centerline{(a)}
}
	\parbox{0.45\textwidth}{
	\includegraphics[width=0.45\textwidth]{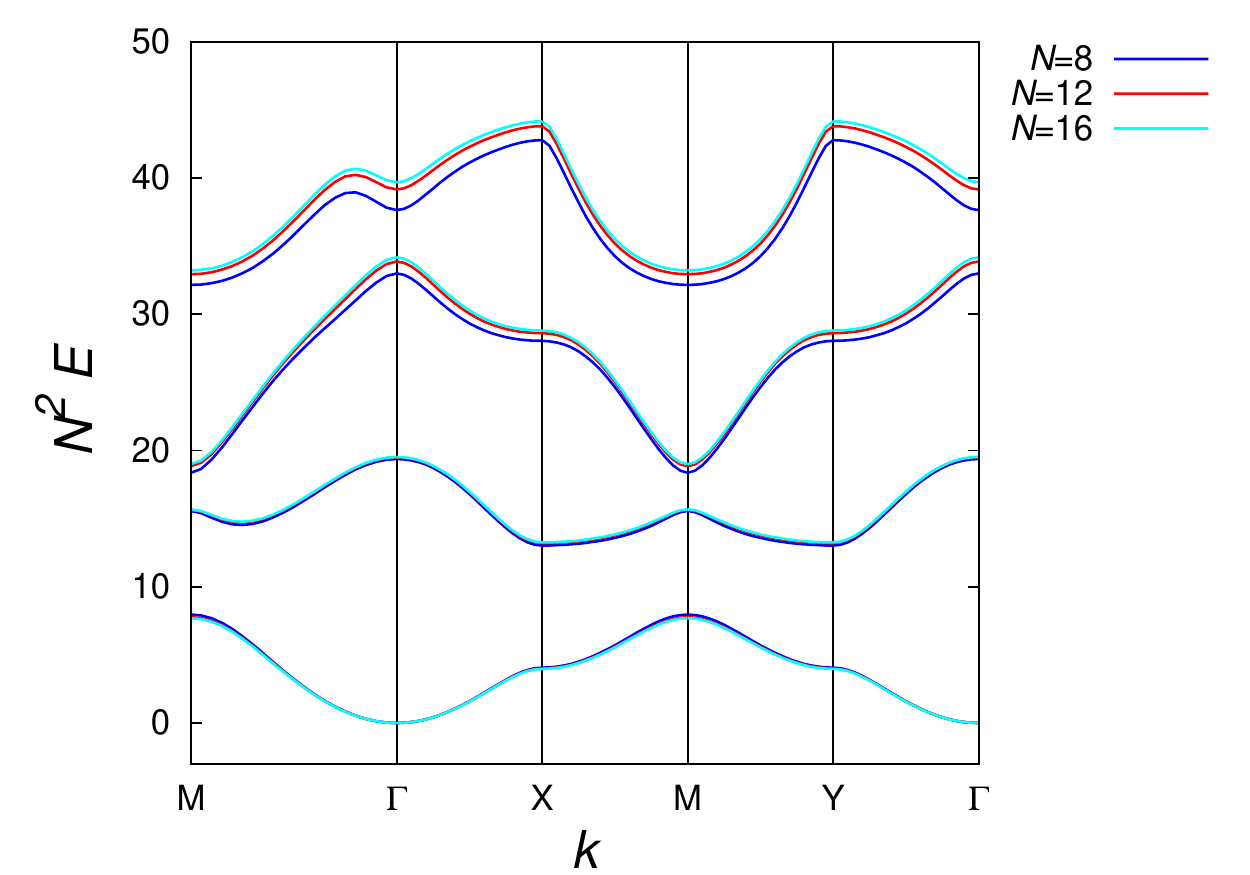}
	\centerline{(b)}
}

\caption{ \label{fig:sE}
		Scaled energy dispersion $N^2 E$ for $N=8,16,32$ under
			  (a) uniform magnetic field ($N=M$);
			  (b) non-uniform magnetic field ($M/N=3/4$).
	}
\end{figure}
\begin{figure}
	\includegraphics[width=0.7\textwidth]{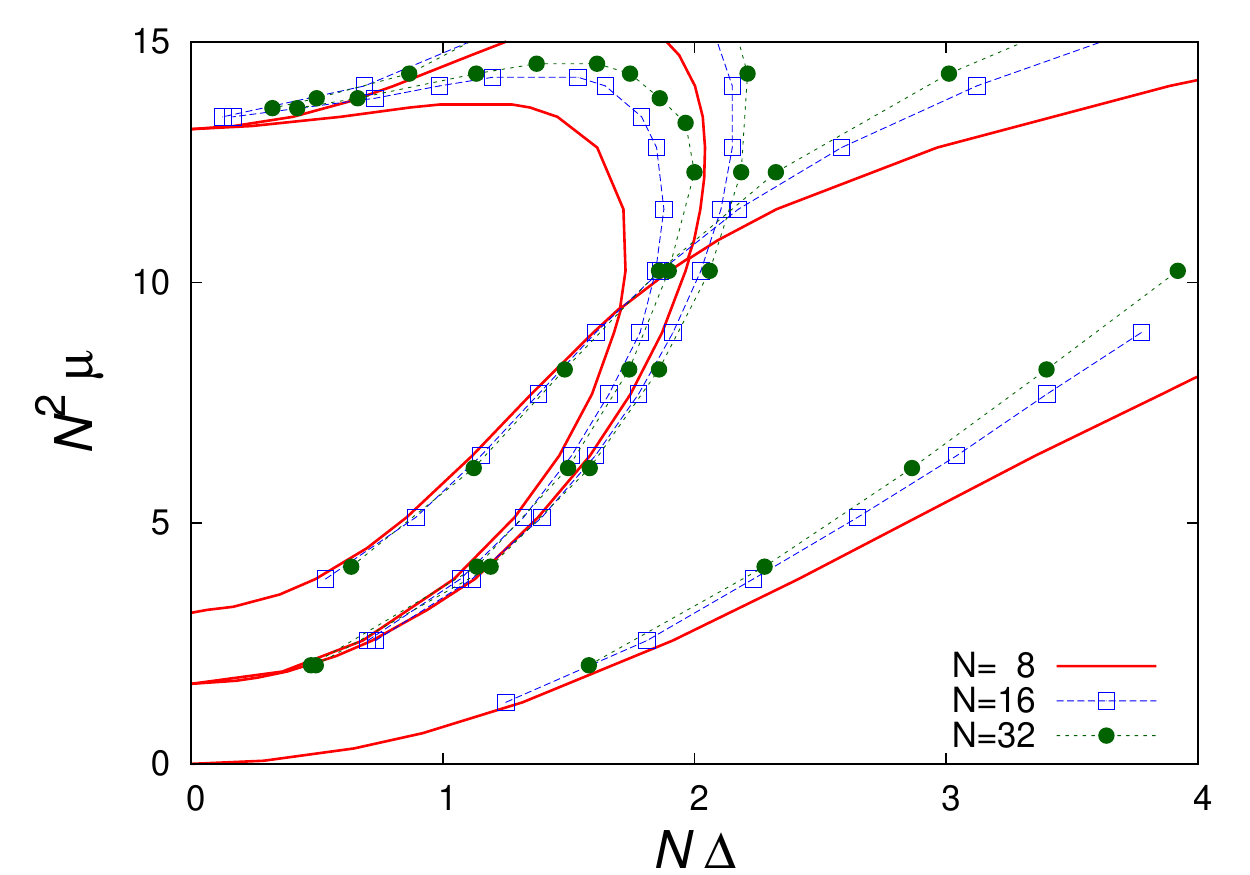}
	\caption{ \label{fig:sPD}
		Phase diagram in the plane of scaled chemical potential $N^2\mu$ and scaled SC pairing parameter $N \Delta$ for $N=8,16,32$ under non-uniform magnetic field
			   ($M/N=3/4$).
	}
\end{figure}

One may suspect that the phase boundaries are also close to the thermodynamic limit. To explicitly demonstrate that, in  Fig.~\ref{fig:sPD} we plot the boundaries for the topological phases for $N=8,16,32$.
It turns out that when plotted as a function of $N \Delta$ as well as $N^2 E$, we obtain a collapse of the phase boundaries for small $\Delta$, which is the region of maximum interest to us.
As $\Delta$ is increased, the phase boundaries for different $N$ begin to deviate from each other, but still capture the thermodynamic phase diagram qualitatively and semiquantitatively.

\section{periodic $h/2e$ fluxoids of a hexagonal shape in triangular lattice}

\begin{figure}
	\parbox{0.35\textwidth}{
	\includegraphics[width=0.35\textwidth]{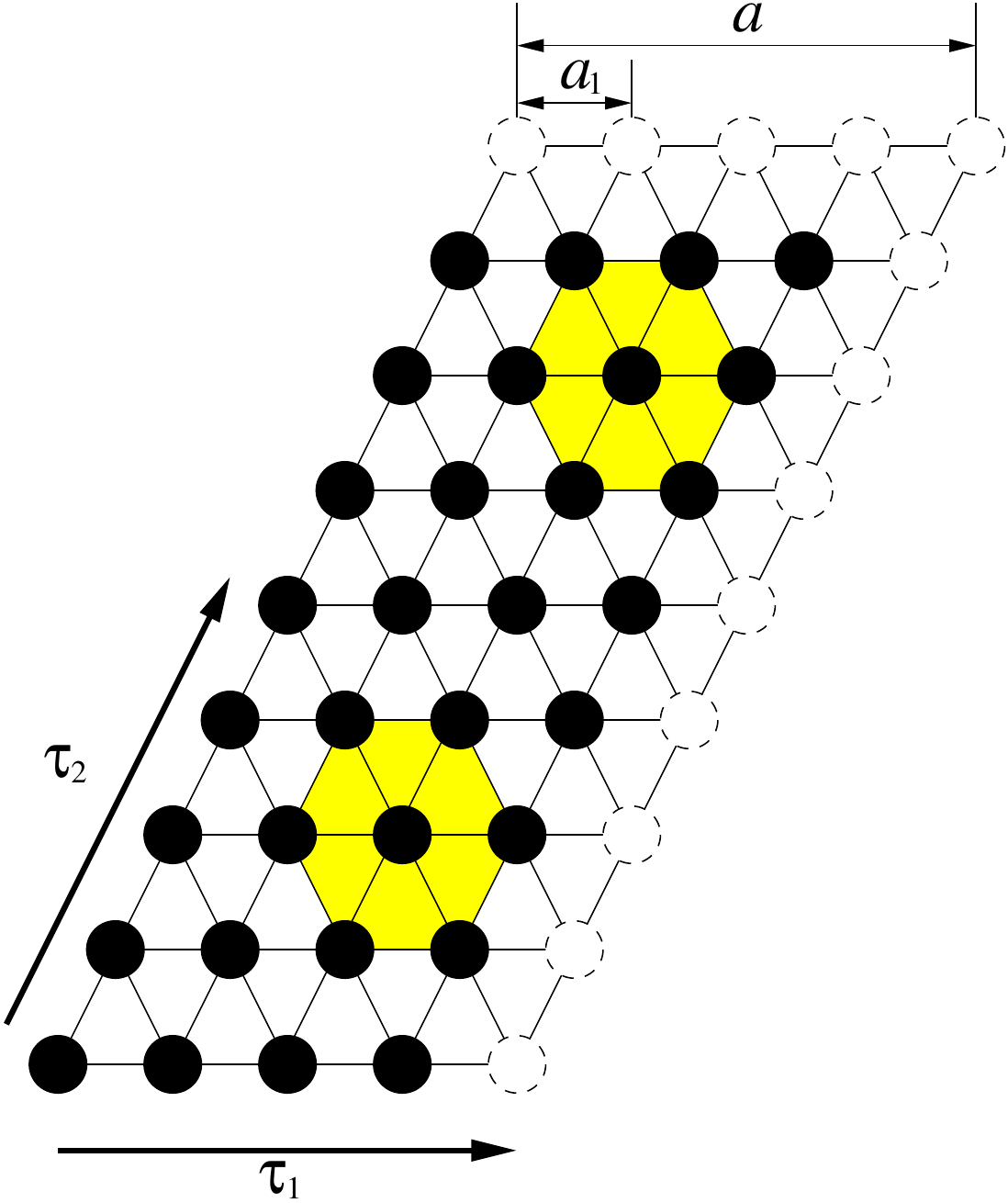}
	\centerline{(a)}
}
	\parbox{0.45\textwidth}{
	\includegraphics[width=0.6\textwidth]{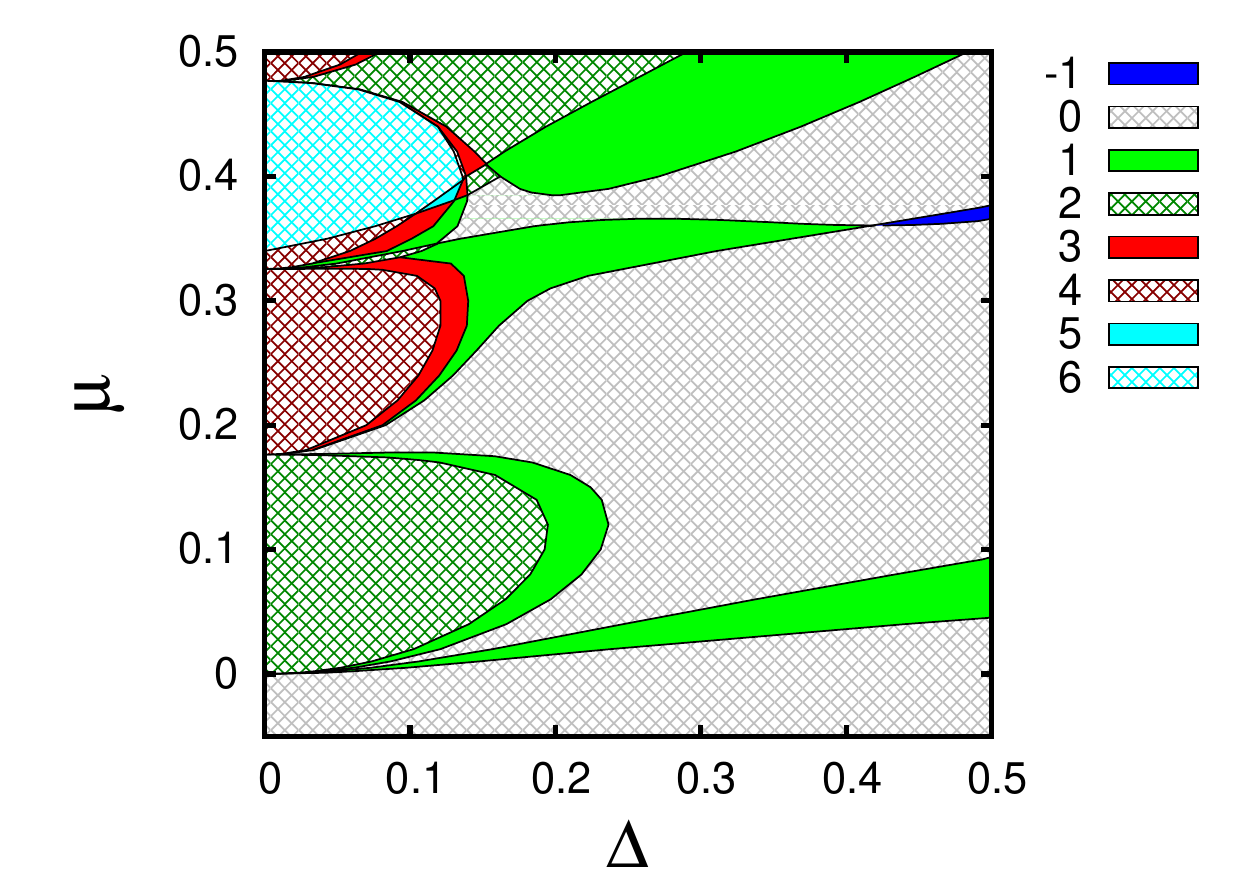}
	\centerline{(b)}
}
\caption{
	\label{fig:h2e}
		(a) Schematic of a unit cell  and
		(b) phase diagram under periodic $h/2e$ fluxoids of hexagonal shape ($N=8,M=6$) in triangular lattice.
		One unit cell is composed of $N\times 2N$ lattice points and two yellow-shaded zones of hexagon shape with $3M^2/2$ triangles contain $h/2e$ fluxoids, respectively. The figure in (a) is an example of $N=4$ and $M=2$.
	}
\end{figure}

We have also performed numerical calculations to obtain the phase diagram for a hexagonal Abrikosov lattice of $h/2e$ fluxoids.
		 We use the same form of full Hamiltonian as in the main text.
		 The lattice points are arranged on the triangular lattice as shown in Fig.~\ref{fig:h2e}(a), and the nearest-neighbor vectors
		 $\bdelta_{1,2,3} = (1,0),(1/2,\sqrt{3}/2),(-1/2,\sqrt{3}/2) $ and the primitive lattice vectors can be chosen as ${\bf a}_1=\bdelta_1=(1,0)$ and ${\bf a}_2=\bdelta_2=(1/2,\sqrt{3}/2)$ on the cartesian coordinate. In an Abrikosov lattice of $h/2e$ fluxoids,
		 one (magnetic) unit cell contains two $h/2e$ fluxoids (see Fig.~\ref{fig:h2e}(a)). Similar to square lattice, the lattice site in this triangular lattice can be labelled as $\vec{r}_{\bm{i}=(i_1,i_2)} = \vec{R}_{\bm{m}=(m_1,m_2)} + \vec{\tilde{r}}_{\bm{l}=(l_1,l_2)}$
for $ (i_1,i_2) = (N m_1 + l_1,2N m_2 + l_2)$ with
$0\le l_{1} < N$, $0\le l_{2} < 2N$, $0\le m_{1/2} < L $, and $L a$ and $2L a$ being the linear dimensions of the system along the ${\bf a}_1$ and ${\bf a}_2$ directions, respectively.
On the basis of ${\bf a}_1$ and ${\bf a}_2$, the lattice vector $\vec{r}_{\bm{i}=(i_x,i_y)}$ can be expressed as $\vec{r}_{\bm{i}=(i_1,i_2)}=i_1{\bf a}_1 + i_2 {\bf a}_2$.

		 The resulting phase diagram in Fig.~\ref{fig:h2e}(b) reveals that there is no topological superconductor in the limit of $\Delta \rightarrow 0$.
		 The TSC shows up for rather large $\Delta$ in the limited region.
		 As in the square lattice, the TSC and normal superconductors are separated by the gap closing at the symmetry points, and two-fold gap-closing boundaries are also existing.

\begin{figure}
	\parbox{0.35\textwidth}{
	\includegraphics[width=0.25\textwidth]{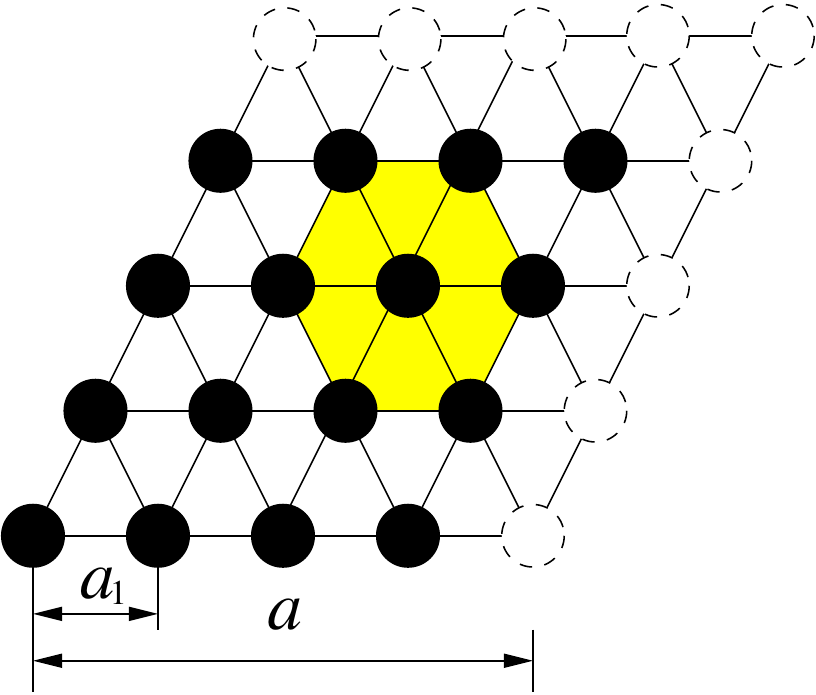}
	\centerline{(a)}
}
	\parbox{0.45\textwidth}{
	\includegraphics[width=0.6\textwidth]{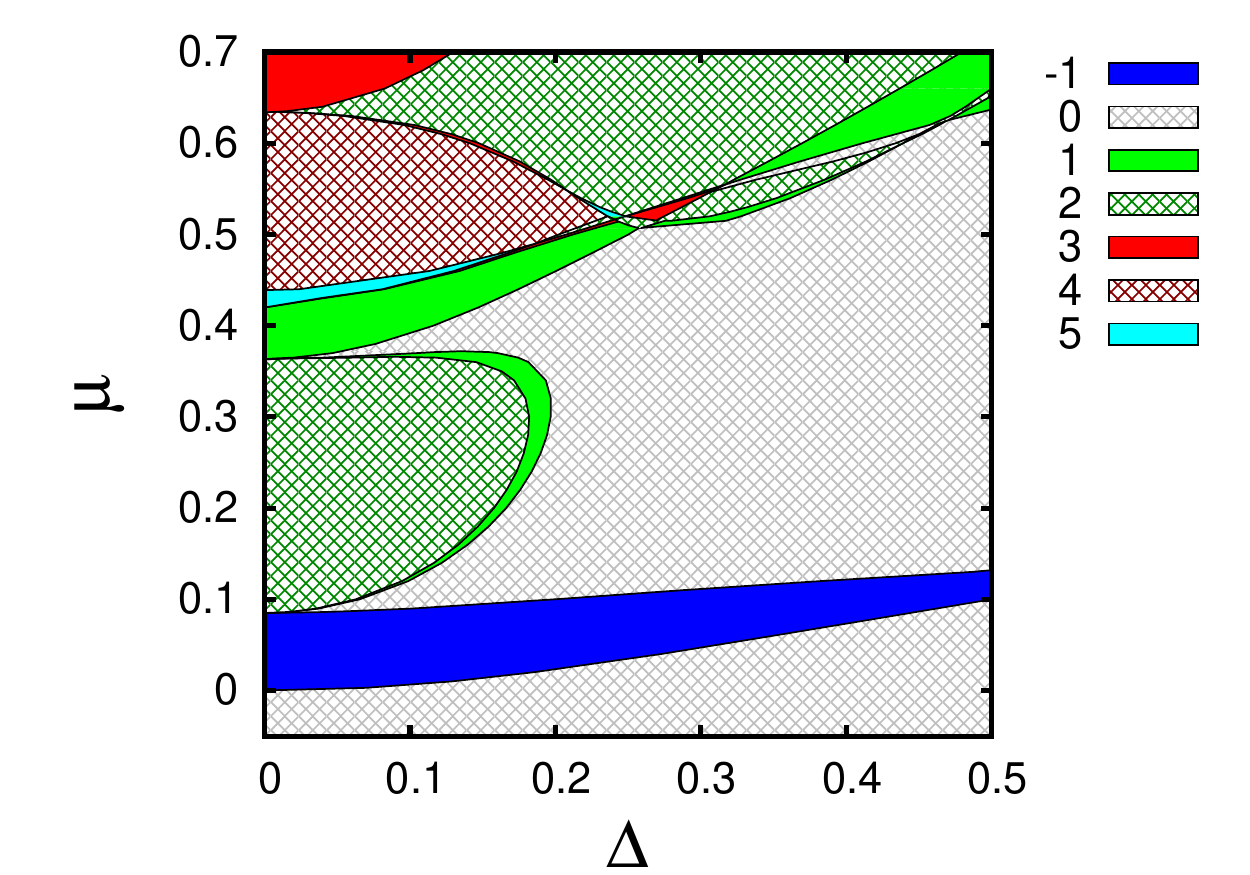}
	\centerline{(b)}
}
	\caption{
		\label{fig:he}
		(a) Schematic of a unit cell  and
		(b) phase diagram under periodic $h/e$ fluxoids of hexagonal shape ($N=8,M=6$) in triangular lattice.
		One unit cell is composed of $N\times N$ lattice points and a yellow-shaded zone of hexagon shape with $3M^2/2$ triangles contain $h/e$ fluxoids. The figure in (a) is an example of $N=4$ and $M=2$.
	}
\end{figure}

The absence of TSC phase in the $\Delta\rightarrow 0$ limit originates from an additional fundamental symmetry of $h/2e$ fluxoid lattice, which we analyze below. The Hamiltonian $\mathcal{H}$ still takes the form of Eq. (1) in the main text, but with ${\bf j}, \bdelta$ defined on the triangle lattice. For convenience, we separate $\mathcal{H}$ into two parts, $\mathcal{H}=\mathcal{H}_0+\mathcal{H}_{\Delta}$ where $\mathcal{H}_0$ includes the terms with the hopping parameter $t$ and the chemical potential $\mu$ and $\mathcal{H}_{\Delta}$ includes the terms with pairing parameter $\Delta$. We first focus on the single-particle Hamiltonian $\mathcal{H}_0$ and note that there are three terms in $\mathcal{H}_0$, which describes the hopping in three directions, described by $\bdelta_1=(1,0)$, $\bdelta_2=(0,1)$ and $\bdelta_3=(-1,1)$ on the basis of ${\bf a}_1$ and ${\bf a}_2$. This is different from the square lattice which only has hopping along two directions. The Hamiltonian $\mathcal{H}_0$ is invariant under the following magnetic translation operators
\begin{eqnarray}
\hat{T}_1=\sum_{\bj}c^{\dag}_{\bj+\btau_1}c_{\bj} e^{i\sum_{\bl_1}\chi_{\bj+{\bl}_1+\bdelta_1,\bj+{\bl}_1}},\nonumber\\
\hat{T}_2=\sum_{\bj}c^{\dag}_{\bj+\btau_2}c_{\bj} e^{i\sum_{\bl_2}\chi_{\bj+{\bl}_2+\bdelta_2,\bj+{\bl}_2}},
\label{eq:Magnetic_translation}
\end{eqnarray}
where $\bj=(j_1,j_2)$ with the integers $j_1, j_2$, $\btau_{1}=(N,0)$, $\btau_{2}=(0,N)$, $\bdelta_1=(1,0)$, $\bdelta_2=(0,1)$, $\bl_1=(l_1,0)$ with the integer $0\le l_1<N-1$, and $\bl_2=(0,l_2)$ with $0\le l_2<N-1$. Here all the vectors are defined on the basis of ${\bf a}_1$ and ${\bf a}_2$. It should be noted that $\btau_1$ is the translation of one magnetic unit cell along the ${\bf a}_1$ direction while $\btau_2$ is the translation over {\it half} magnetic unit cell along the ${\bf a}_2$ direction (see Fig.~ \ref{fig:h2e}(a))
The phase factor field $\chi_{\bj+\bdelta,\bj}$ is defined on each bond of the lattice and can be determined by the vector potential field $A_{\bj+\bdelta,\bj}$ as
\begin{eqnarray}
&&\nabla_1\chi_{\bj+\bdelta_1,\bj}=\nabla_1 A_{\bj+\bdelta_1,\bj},\label{eq:chi_A_1}\\
&&\nabla_2\chi_{\bj+\bdelta_1,\bj}=\nabla_1 A_{\bj+\bdelta_2,\bj}=\nabla_2 A_{\bj+\bdelta_1,\bj}+2\pi \phi_{\bj},\label{eq:chi_A_2}\\
&&\nabla_2\chi_{\bj+\bdelta_2,\bj}=\nabla_2 A_{\bj+\bdelta_2,\bj},\label{eq:chi_A_3}\\
&&\nabla_1\chi_{\bj+\bdelta_2,\bj}=\nabla_2 A_{\bj+\bdelta_1,\bj}=\nabla_1 A_{\bj+\bdelta_2,\bj}-2\pi \phi_{\bj}. \label{eq:chi_A_4}
\end{eqnarray}
Here the discrete differential operator $\nabla$ is defined as $\nabla_a f_{\bj+\bdelta_b,\bj}=f_{\bj+\bdelta_b+\bdelta_a,\bj+\bdelta_a}-f_{\bj+\bdelta_b,\bj}$ with $a,b=1,2$ and $f=\chi, A$, and $2\pi \phi_{\bj}=\nabla_1 A_{\bj+\bdelta_2,\bj}-\nabla_2 A_{\bj+\bdelta_1,\bj}=A_{\bj+\bdelta_2+\bdelta_1,\bj+\bdelta_1}-A_{\bj+\bdelta_2,\bj}
-A_{\bj+\bdelta_1+\bdelta_2,\bj+\bdelta_2}+A_{\bj+\bdelta_1,\bj}$ labels the flux in the plaquette of the rhomboid formed by four sites $\bj$, $\bj+\bdelta_1$, $\bj+\bdelta_2$ and $\bj+\bdelta_1+\bdelta_2$.
Here the flux $\phi_{\bj}$ satisfies the periodic conditions $\phi_{\bj}=\phi_{\bj+\btau_a}$ with $a=1,2$ and $\sum_{\bj}\phi_{\bj}=1/2$ where the summation over $\bj=(j_1,j_2)$ is within the range $0\le j_{1,2}<N$.
The phase factor $\chi_{\bj+\bdelta_1,\bj}$ ($\chi_{\bj+\bdelta_2,\bj}$) determined by \ref{eq:chi_A_1} and \ref{eq:chi_A_2} (\ref{eq:chi_A_3} and \ref{eq:chi_A_4}) will make the operator $\hat{T}_1$ ($\hat{T}_2$) commutate with the hopping terms along the $\bdelta_1$ and $\bdelta_2$ directions in $\mathcal{H}_0$, which is similar to the case of a square lattice under magnetic fields \cite{Bernevig13}. For the hopping along the $\bdelta_3$ direction, one can still show its commutation with $\hat{T}_{a}$ due to the periodic condition $\phi_{\bj}=\phi_{\bj+\btau_a}$ after some length derivation. Therefore, $\hat{T}_{a}$ defines the magnetic translation ($[\hat{T}_a,\mathcal{H}_0]=0, a=1,2$) for our system. In addition, since the rhomboid formed by two vectors $\btau_1$ and $\btau_2$ encloses half flux quantum for our $h/2e$ fluxoid lattice, we find the anti-commutation relation $\{\hat{T}_1,\hat{T}_2\}=0$ within the single-particle Hilbert space. These commutation or anti-commutation relations within the operation sets of $\hat{T}_1$, $\hat{T}_2$ and $\mathcal{H}_0$ determine the energy spectrum of the Abrikosov lattice system. 

The eigen-state $|\bk\rangle$ of the Hamiltonian $\mathcal{H}_0$ can be chosen as $\mathcal{H}_0|\bk\rangle=E(\bk)|\bk\rangle$, $\hat{T}_1|\bk\rangle=e^{i k_1}|\bk\rangle$ and $\hat{T}^2_1|\bk\rangle=e^{i 2k_2}|\bk\rangle$ with $\bk=(k_1,k_2)$ since $\mathcal{H}_0$, $\hat{T}_1$ and $\hat{T}^2_2$ all commutate with each other. Here $k_1$ and $k_2$ are continuous variables in the range $-\pi\le k_1<\pi$ and $-\pi/2\le k_2<\pi/2$. Let us consider two eigen-states $|\bk\rangle$ and $\hat{T}_2|\bk\rangle$ and it is clear that these two eigen-states share the same eigen-energy since $[\mathcal{H}_0,\hat{T}_2]=0$. Due to the anti-commutation between $\hat{T}_1$ and $\hat{T}_2$, we find $\hat{T}_1\hat{T}_2|\bk\rangle=-\hat{T}_2\hat{T}_1|\bk\rangle=-e^{ik_1}\hat{T}_2|\bk\rangle
=e^{i(k_1+\pi)}\hat{T}_2|\bk\rangle$. This suggests that $\hat{T}_2|\bk\rangle\sim |\bk+(\pi,0)\rangle$ and all the eigen-states at $\bk$ and $\bk+(\pi,0)$ share the same energy spectrum. Based on this conclusion, one can see that Dirac type of topological phase transition must occur in pairs at $\bk$ and $\bk+(\pi,0)$ in the entire Brillouin zone and thus the SC Chern number is always changed by $\pm 2$ in the $\Delta\rightarrow 0$ limit. Thus, there is no TSC phase in this limit. This argument is not valid for a finite $\Delta$, because the p-wave pairing term $\mathcal{H}_{\Delta}$ that we choose for the calculations is {\it not} gauge invariant under magnetic translations $\hat{T}_{1,2}$. Therefore, the TSC phase can still be present at a finite $\Delta$.

\section{periodic $h/e$ fluxoids of a hexagonal shape in triangular lattice}

We finally show the phase diagram for topological superconductors in the presence of a hexagonal lattice of $h/e$ fluxoids.
The lattice points are arranged on the triangular lattice as shown in Fig.~\ref{fig:he}(a), and the nearest-neighbor vectors are given by
		 $\bdelta = (1,0),(1/2,\sqrt{3}/2),(-1/2,\sqrt{3}/2) $.
		 Each unit cell contains a single $h/e$ fluxoid.
		 This geometry also produces topological superconductor in the limit of $\Delta \rightarrow 0$, as demonstrated in Fig.~\ref{fig:he}(b).
		 Some TSC intervenes between the adjacent normal quantum Hall phases.
		 This is similar to the geometry of periodic square fluxes.

Compared to the periodic $h/2e$ fluxoid lattice, we notice that the magnetic translation symmetry and the lattice translation in the periodic $h/e$ fluxoid lattice coincide with each other and thus we do not have any additional symmetry leading to degeneracy. As a result, TSC phase is allowed in the limit $\Delta\rightarrow 0$.

\end{widetext}

\bibliography{biblio_fqhe}

\begin{thebibliography}{49}
\expandafter\ifx\csname natexlab\endcsname\relax\def\natexlab#1{#1}\fi
\expandafter\ifx\csname bibnamefont\endcsname\relax
  \def\bibnamefont#1{#1}\fi
\expandafter\ifx\csname bibfnamefont\endcsname\relax
  \def\bibfnamefont#1{#1}\fi
\expandafter\ifx\csname citenamefont\endcsname\relax
  \def\citenamefont#1{#1}\fi
\expandafter\ifx\csname url\endcsname\relax
  \def\url#1{\texttt{#1}}\fi
\expandafter\ifx\csname urlprefix\endcsname\relax\def\urlprefix{URL }\fi
\providecommand{\bibinfo}[2]{#2}
\providecommand{\eprint}[2][]{\url{#2}}

\bibitem[{\citenamefont{Stern}(2008)}]{stern2008anyons}
\bibinfo{author}{\bibfnamefont{A.}~\bibnamefont{Stern}},
  \bibinfo{journal}{Annals of Physics} \textbf{\bibinfo{volume}{323}},
  \bibinfo{pages}{204} (\bibinfo{year}{2008}).

\bibitem[{\citenamefont{Nayak et~al.}(2008)\citenamefont{Nayak, Simon, Stern,
  Freedman, and Das~Sarma}}]{Nayak08}
\bibinfo{author}{\bibfnamefont{C.}~\bibnamefont{Nayak}},
  \bibinfo{author}{\bibfnamefont{S.~H.} \bibnamefont{Simon}},
  \bibinfo{author}{\bibfnamefont{A.}~\bibnamefont{Stern}},
  \bibinfo{author}{\bibfnamefont{M.}~\bibnamefont{Freedman}}, \bibnamefont{and}
  \bibinfo{author}{\bibfnamefont{S.}~\bibnamefont{Das~Sarma}},
  \bibinfo{journal}{Rev. Mod. Phys.} \textbf{\bibinfo{volume}{80}},
  \bibinfo{pages}{1083} (\bibinfo{year}{2008}),
  \urlprefix\url{http://link.aps.org/doi/10.1103/RevModPhys.80.1083}.

\bibitem[{\citenamefont{Alicea}(2012)}]{alicea_rpp_2012}
\bibinfo{author}{\bibfnamefont{J.}~\bibnamefont{Alicea}},
  \bibinfo{journal}{Reports on Progress in Physics}
  \textbf{\bibinfo{volume}{75}}, \bibinfo{pages}{076501}
  (\bibinfo{year}{2012}).

\bibitem[{\citenamefont{Beenakker}(2013)}]{beenakker2013search}
\bibinfo{author}{\bibfnamefont{C.}~\bibnamefont{Beenakker}}
  (\bibinfo{year}{2013}).

\bibitem[{\citenamefont{Kitaev}(2001)}]{kitaev2001unpaired}
\bibinfo{author}{\bibfnamefont{A.~Y.} \bibnamefont{Kitaev}},
  \bibinfo{journal}{Physics-Uspekhi} \textbf{\bibinfo{volume}{44}},
  \bibinfo{pages}{131} (\bibinfo{year}{2001}).

\bibitem[{\citenamefont{Kitaev}(2003)}]{Kitaev03}
\bibinfo{author}{\bibfnamefont{A.}~\bibnamefont{Kitaev}},
  \bibinfo{journal}{Annals of Physics} \textbf{\bibinfo{volume}{303}},
  \bibinfo{pages}{2 } (\bibinfo{year}{2003}), ISSN \bibinfo{issn}{0003-4916},
  \urlprefix\url{http://www.sciencedirect.com/science/article/pii/S0003491602000180}.

\bibitem[{\citenamefont{Sarma et~al.}(2015)\citenamefont{Sarma, Freedman, and
  Nayak}}]{sarma2015majorana}
\bibinfo{author}{\bibfnamefont{S.~D.} \bibnamefont{Sarma}},
  \bibinfo{author}{\bibfnamefont{M.}~\bibnamefont{Freedman}}, \bibnamefont{and}
  \bibinfo{author}{\bibfnamefont{C.}~\bibnamefont{Nayak}},
  \bibinfo{journal}{npj Quantum Information} \textbf{\bibinfo{volume}{1}},
  \bibinfo{pages}{15001} (\bibinfo{year}{2015}).

\bibitem[{\citenamefont{Alicea and Stern}(2015)}]{alicea2015designer}
\bibinfo{author}{\bibfnamefont{J.}~\bibnamefont{Alicea}} \bibnamefont{and}
  \bibinfo{author}{\bibfnamefont{A.}~\bibnamefont{Stern}},
  \bibinfo{journal}{Physica Scripta} \textbf{\bibinfo{volume}{2015}},
  \bibinfo{pages}{014006} (\bibinfo{year}{2015}).

\bibitem[{\citenamefont{Read and Green}(2000)}]{Read00}
\bibinfo{author}{\bibfnamefont{N.}~\bibnamefont{Read}} \bibnamefont{and}
  \bibinfo{author}{\bibfnamefont{D.}~\bibnamefont{Green}},
  \bibinfo{journal}{Phys. Rev. B} \textbf{\bibinfo{volume}{61}},
  \bibinfo{pages}{10267} (\bibinfo{year}{2000}),
  \urlprefix\url{http://link.aps.org/doi/10.1103/PhysRevB.61.10267}.

\bibitem[{\citenamefont{Rice and Sigrist}(1995)}]{rice1995sr2ruo4}
\bibinfo{author}{\bibfnamefont{T.}~\bibnamefont{Rice}} \bibnamefont{and}
  \bibinfo{author}{\bibfnamefont{M.}~\bibnamefont{Sigrist}},
  \bibinfo{journal}{Journal of Physics: Condensed Matter}
  \textbf{\bibinfo{volume}{7}}, \bibinfo{pages}{L643} (\bibinfo{year}{1995}).

\bibitem[{\citenamefont{Mourik et~al.}(2012)\citenamefont{Mourik, Zuo, Frolov,
  Plissard, Bakkers, and Kouwenhoven}}]{mourik2012signatures}
\bibinfo{author}{\bibfnamefont{V.}~\bibnamefont{Mourik}},
  \bibinfo{author}{\bibfnamefont{K.}~\bibnamefont{Zuo}},
  \bibinfo{author}{\bibfnamefont{S.~M.} \bibnamefont{Frolov}},
  \bibinfo{author}{\bibfnamefont{S.}~\bibnamefont{Plissard}},
  \bibinfo{author}{\bibfnamefont{E.~P.} \bibnamefont{Bakkers}},
  \bibnamefont{and} \bibinfo{author}{\bibfnamefont{L.~P.}
  \bibnamefont{Kouwenhoven}}, \bibinfo{journal}{Science}
  \textbf{\bibinfo{volume}{336}}, \bibinfo{pages}{1003} (\bibinfo{year}{2012}).

\bibitem[{\citenamefont{Das et~al.}(2012)\citenamefont{Das, Ronen, Most, Oreg,
  Heiblum, and Shtrikman}}]{das2012zero}
\bibinfo{author}{\bibfnamefont{A.}~\bibnamefont{Das}},
  \bibinfo{author}{\bibfnamefont{Y.}~\bibnamefont{Ronen}},
  \bibinfo{author}{\bibfnamefont{Y.}~\bibnamefont{Most}},
  \bibinfo{author}{\bibfnamefont{Y.}~\bibnamefont{Oreg}},
  \bibinfo{author}{\bibfnamefont{M.}~\bibnamefont{Heiblum}}, \bibnamefont{and}
  \bibinfo{author}{\bibfnamefont{H.}~\bibnamefont{Shtrikman}},
  \bibinfo{journal}{Nature Physics} \textbf{\bibinfo{volume}{8}},
  \bibinfo{pages}{887} (\bibinfo{year}{2012}).

\bibitem[{\citenamefont{Rokhinson et~al.}(2012)\citenamefont{Rokhinson, Liu,
  and Furdyna}}]{rokhinson2012fractional}
\bibinfo{author}{\bibfnamefont{L.~P.} \bibnamefont{Rokhinson}},
  \bibinfo{author}{\bibfnamefont{X.}~\bibnamefont{Liu}}, \bibnamefont{and}
  \bibinfo{author}{\bibfnamefont{J.~K.} \bibnamefont{Furdyna}},
  \bibinfo{journal}{Nature Physics} \textbf{\bibinfo{volume}{8}},
  \bibinfo{pages}{795} (\bibinfo{year}{2012}).

\bibitem[{\citenamefont{Lutchyn et~al.}(2010)\citenamefont{Lutchyn, Sau, and
  Sarma}}]{lutchyn2010majorana}
\bibinfo{author}{\bibfnamefont{R.~M.} \bibnamefont{Lutchyn}},
  \bibinfo{author}{\bibfnamefont{J.~D.} \bibnamefont{Sau}}, \bibnamefont{and}
  \bibinfo{author}{\bibfnamefont{S.~D.} \bibnamefont{Sarma}},
  \bibinfo{journal}{Physical review letters} \textbf{\bibinfo{volume}{105}},
  \bibinfo{pages}{077001} (\bibinfo{year}{2010}).

\bibitem[{\citenamefont{Sau et~al.}(2010)\citenamefont{Sau, Lutchyn, Tewari,
  and Sarma}}]{sau2010generic}
\bibinfo{author}{\bibfnamefont{J.~D.} \bibnamefont{Sau}},
  \bibinfo{author}{\bibfnamefont{R.~M.} \bibnamefont{Lutchyn}},
  \bibinfo{author}{\bibfnamefont{S.}~\bibnamefont{Tewari}}, \bibnamefont{and}
  \bibinfo{author}{\bibfnamefont{S.~D.} \bibnamefont{Sarma}},
  \bibinfo{journal}{Physical review letters} \textbf{\bibinfo{volume}{104}},
  \bibinfo{pages}{040502} (\bibinfo{year}{2010}).

\bibitem[{\citenamefont{Alicea}(2010)}]{alicea2010majorana}
\bibinfo{author}{\bibfnamefont{J.}~\bibnamefont{Alicea}},
  \bibinfo{journal}{Physical Review B} \textbf{\bibinfo{volume}{81}},
  \bibinfo{pages}{125318} (\bibinfo{year}{2010}).

\bibitem[{\citenamefont{Alicea et~al.}(2011)\citenamefont{Alicea, Oreg, Refael,
  Von~Oppen, and Fisher}}]{alicea2011non}
\bibinfo{author}{\bibfnamefont{J.}~\bibnamefont{Alicea}},
  \bibinfo{author}{\bibfnamefont{Y.}~\bibnamefont{Oreg}},
  \bibinfo{author}{\bibfnamefont{G.}~\bibnamefont{Refael}},
  \bibinfo{author}{\bibfnamefont{F.}~\bibnamefont{Von~Oppen}},
  \bibnamefont{and} \bibinfo{author}{\bibfnamefont{M.~P.}
  \bibnamefont{Fisher}}, \bibinfo{journal}{Nature Physics}
  \textbf{\bibinfo{volume}{7}}, \bibinfo{pages}{412} (\bibinfo{year}{2011}).

\bibitem[{\citenamefont{Nadj-Perge et~al.}(2014)\citenamefont{Nadj-Perge,
  Drozdov, Li, Chen, Jeon, Seo, MacDonald, Bernevig, and
  Yazdani}}]{nadj2014observation}
\bibinfo{author}{\bibfnamefont{S.}~\bibnamefont{Nadj-Perge}},
  \bibinfo{author}{\bibfnamefont{I.~K.} \bibnamefont{Drozdov}},
  \bibinfo{author}{\bibfnamefont{J.}~\bibnamefont{Li}},
  \bibinfo{author}{\bibfnamefont{H.}~\bibnamefont{Chen}},
  \bibinfo{author}{\bibfnamefont{S.}~\bibnamefont{Jeon}},
  \bibinfo{author}{\bibfnamefont{J.}~\bibnamefont{Seo}},
  \bibinfo{author}{\bibfnamefont{A.~H.} \bibnamefont{MacDonald}},
  \bibinfo{author}{\bibfnamefont{B.~A.} \bibnamefont{Bernevig}},
  \bibnamefont{and} \bibinfo{author}{\bibfnamefont{A.}~\bibnamefont{Yazdani}},
  \bibinfo{journal}{Science} \textbf{\bibinfo{volume}{346}},
  \bibinfo{pages}{602} (\bibinfo{year}{2014}).

\bibitem[{\citenamefont{Fu and Kane}(2008)}]{fu2008superconducting}
\bibinfo{author}{\bibfnamefont{L.}~\bibnamefont{Fu}} \bibnamefont{and}
  \bibinfo{author}{\bibfnamefont{C.~L.} \bibnamefont{Kane}},
  \bibinfo{journal}{Physical review letters} \textbf{\bibinfo{volume}{100}},
  \bibinfo{pages}{096407} (\bibinfo{year}{2008}).

\bibitem[{\citenamefont{Xu et~al.}(2015)\citenamefont{Xu, Wang, Liu, Ge, Yang,
  Liu, Xu, Guan, Gao, Qian et~al.}}]{xu2015experimental}
\bibinfo{author}{\bibfnamefont{J.-P.} \bibnamefont{Xu}},
  \bibinfo{author}{\bibfnamefont{M.-X.} \bibnamefont{Wang}},
  \bibinfo{author}{\bibfnamefont{Z.~L.} \bibnamefont{Liu}},
  \bibinfo{author}{\bibfnamefont{J.-F.} \bibnamefont{Ge}},
  \bibinfo{author}{\bibfnamefont{X.}~\bibnamefont{Yang}},
  \bibinfo{author}{\bibfnamefont{C.}~\bibnamefont{Liu}},
  \bibinfo{author}{\bibfnamefont{Z.~A.} \bibnamefont{Xu}},
  \bibinfo{author}{\bibfnamefont{D.}~\bibnamefont{Guan}},
  \bibinfo{author}{\bibfnamefont{C.~L.} \bibnamefont{Gao}},
  \bibinfo{author}{\bibfnamefont{D.}~\bibnamefont{Qian}}, \bibnamefont{et~al.},
  \bibinfo{journal}{Physical review letters} \textbf{\bibinfo{volume}{114}},
  \bibinfo{pages}{017001} (\bibinfo{year}{2015}).

\bibitem[{\citenamefont{Wang et~al.}(2012)\citenamefont{Wang, Liu, Xu, Yang,
  Miao, Yao, Gao, Shen, Ma, Chen et~al.}}]{wang2012coexistence}
\bibinfo{author}{\bibfnamefont{M.-X.} \bibnamefont{Wang}},
  \bibinfo{author}{\bibfnamefont{C.}~\bibnamefont{Liu}},
  \bibinfo{author}{\bibfnamefont{J.-P.} \bibnamefont{Xu}},
  \bibinfo{author}{\bibfnamefont{F.}~\bibnamefont{Yang}},
  \bibinfo{author}{\bibfnamefont{L.}~\bibnamefont{Miao}},
  \bibinfo{author}{\bibfnamefont{M.-Y.} \bibnamefont{Yao}},
  \bibinfo{author}{\bibfnamefont{C.}~\bibnamefont{Gao}},
  \bibinfo{author}{\bibfnamefont{C.}~\bibnamefont{Shen}},
  \bibinfo{author}{\bibfnamefont{X.}~\bibnamefont{Ma}},
  \bibinfo{author}{\bibfnamefont{X.}~\bibnamefont{Chen}}, \bibnamefont{et~al.},
  \bibinfo{journal}{Science} \textbf{\bibinfo{volume}{336}},
  \bibinfo{pages}{52} (\bibinfo{year}{2012}).

\bibitem[{\citenamefont{Yin et~al.}(2015)\citenamefont{Yin, Wu, Wang, Ye, Gong,
  Hou, Shan, Li, Liang, Wu et~al.}}]{yin2015observation}
\bibinfo{author}{\bibfnamefont{J.}~\bibnamefont{Yin}},
  \bibinfo{author}{\bibfnamefont{Z.}~\bibnamefont{Wu}},
  \bibinfo{author}{\bibfnamefont{J.}~\bibnamefont{Wang}},
  \bibinfo{author}{\bibfnamefont{Z.}~\bibnamefont{Ye}},
  \bibinfo{author}{\bibfnamefont{J.}~\bibnamefont{Gong}},
  \bibinfo{author}{\bibfnamefont{X.}~\bibnamefont{Hou}},
  \bibinfo{author}{\bibfnamefont{L.}~\bibnamefont{Shan}},
  \bibinfo{author}{\bibfnamefont{A.}~\bibnamefont{Li}},
  \bibinfo{author}{\bibfnamefont{X.}~\bibnamefont{Liang}},
  \bibinfo{author}{\bibfnamefont{X.}~\bibnamefont{Wu}}, \bibnamefont{et~al.},
  \bibinfo{journal}{Nature Physics} \textbf{\bibinfo{volume}{11}},
  \bibinfo{pages}{543} (\bibinfo{year}{2015}).

\bibitem[{\citenamefont{Wu et~al.}(2016)\citenamefont{Wu, Qin, Liang, Fan, and
  Hu}}]{wu2016topological}
\bibinfo{author}{\bibfnamefont{X.}~\bibnamefont{Wu}},
  \bibinfo{author}{\bibfnamefont{S.}~\bibnamefont{Qin}},
  \bibinfo{author}{\bibfnamefont{Y.}~\bibnamefont{Liang}},
  \bibinfo{author}{\bibfnamefont{H.}~\bibnamefont{Fan}}, \bibnamefont{and}
  \bibinfo{author}{\bibfnamefont{J.}~\bibnamefont{Hu}},
  \bibinfo{journal}{Physical Review B} \textbf{\bibinfo{volume}{93}},
  \bibinfo{pages}{115129} (\bibinfo{year}{2016}).

\bibitem[{\citenamefont{Zhang et~al.}(2018)\citenamefont{Zhang, Yaji,
  Hashimoto, Ota, Kondo, Okazaki, Wang, Wen, Gu, Ding
  et~al.}}]{zhang2018observation}
\bibinfo{author}{\bibfnamefont{P.}~\bibnamefont{Zhang}},
  \bibinfo{author}{\bibfnamefont{K.}~\bibnamefont{Yaji}},
  \bibinfo{author}{\bibfnamefont{T.}~\bibnamefont{Hashimoto}},
  \bibinfo{author}{\bibfnamefont{Y.}~\bibnamefont{Ota}},
  \bibinfo{author}{\bibfnamefont{T.}~\bibnamefont{Kondo}},
  \bibinfo{author}{\bibfnamefont{K.}~\bibnamefont{Okazaki}},
  \bibinfo{author}{\bibfnamefont{Z.}~\bibnamefont{Wang}},
  \bibinfo{author}{\bibfnamefont{J.}~\bibnamefont{Wen}},
  \bibinfo{author}{\bibfnamefont{G.}~\bibnamefont{Gu}},
  \bibinfo{author}{\bibfnamefont{H.}~\bibnamefont{Ding}}, \bibnamefont{et~al.},
  \bibinfo{journal}{Science} \textbf{\bibinfo{volume}{360}},
  \bibinfo{pages}{182} (\bibinfo{year}{2018}).

\bibitem[{\citenamefont{Wang et~al.}(2015)\citenamefont{Wang, Zhang, Xu, Zeng,
  Miao, Xu, Qian, Weng, Richard, Fedorov et~al.}}]{wang2015topological}
\bibinfo{author}{\bibfnamefont{Z.}~\bibnamefont{Wang}},
  \bibinfo{author}{\bibfnamefont{P.}~\bibnamefont{Zhang}},
  \bibinfo{author}{\bibfnamefont{G.}~\bibnamefont{Xu}},
  \bibinfo{author}{\bibfnamefont{L.}~\bibnamefont{Zeng}},
  \bibinfo{author}{\bibfnamefont{H.}~\bibnamefont{Miao}},
  \bibinfo{author}{\bibfnamefont{X.}~\bibnamefont{Xu}},
  \bibinfo{author}{\bibfnamefont{T.}~\bibnamefont{Qian}},
  \bibinfo{author}{\bibfnamefont{H.}~\bibnamefont{Weng}},
  \bibinfo{author}{\bibfnamefont{P.}~\bibnamefont{Richard}},
  \bibinfo{author}{\bibfnamefont{A.~V.} \bibnamefont{Fedorov}},
  \bibnamefont{et~al.}, \bibinfo{journal}{Physical Review B}
  \textbf{\bibinfo{volume}{92}}, \bibinfo{pages}{115119}
  (\bibinfo{year}{2015}).

\bibitem[{\citenamefont{Qi et~al.}(2010)\citenamefont{Qi, Hughes, and
  Zhang}}]{qi2010chiral}
\bibinfo{author}{\bibfnamefont{X.-L.} \bibnamefont{Qi}},
  \bibinfo{author}{\bibfnamefont{T.~L.} \bibnamefont{Hughes}},
  \bibnamefont{and} \bibinfo{author}{\bibfnamefont{S.-C.} \bibnamefont{Zhang}},
  \bibinfo{journal}{Physical Review B} \textbf{\bibinfo{volume}{82}},
  \bibinfo{pages}{184516} (\bibinfo{year}{2010}).

\bibitem[{\citenamefont{Chung et~al.}(2011)\citenamefont{Chung, Qi, Maciejko,
  and Zhang}}]{chung2011conductance}
\bibinfo{author}{\bibfnamefont{S.~B.} \bibnamefont{Chung}},
  \bibinfo{author}{\bibfnamefont{X.-L.} \bibnamefont{Qi}},
  \bibinfo{author}{\bibfnamefont{J.}~\bibnamefont{Maciejko}}, \bibnamefont{and}
  \bibinfo{author}{\bibfnamefont{S.-C.} \bibnamefont{Zhang}},
  \bibinfo{journal}{Physical Review B} \textbf{\bibinfo{volume}{83}},
  \bibinfo{pages}{100512} (\bibinfo{year}{2011}).

\bibitem[{\citenamefont{He et~al.}(2017)\citenamefont{He, Pan, Stern, Burks,
  Che, Yin, Wang, Lian, Zhou, Choi et~al.}}]{he2017chiral}
\bibinfo{author}{\bibfnamefont{Q.~L.} \bibnamefont{He}},
  \bibinfo{author}{\bibfnamefont{L.}~\bibnamefont{Pan}},
  \bibinfo{author}{\bibfnamefont{A.~L.} \bibnamefont{Stern}},
  \bibinfo{author}{\bibfnamefont{E.~C.} \bibnamefont{Burks}},
  \bibinfo{author}{\bibfnamefont{X.}~\bibnamefont{Che}},
  \bibinfo{author}{\bibfnamefont{G.}~\bibnamefont{Yin}},
  \bibinfo{author}{\bibfnamefont{J.}~\bibnamefont{Wang}},
  \bibinfo{author}{\bibfnamefont{B.}~\bibnamefont{Lian}},
  \bibinfo{author}{\bibfnamefont{Q.}~\bibnamefont{Zhou}},
  \bibinfo{author}{\bibfnamefont{E.~S.} \bibnamefont{Choi}},
  \bibnamefont{et~al.}, \bibinfo{journal}{Science}
  \textbf{\bibinfo{volume}{357}}, \bibinfo{pages}{294} (\bibinfo{year}{2017}).

\bibitem[{\citenamefont{Lian et~al.}(2018)\citenamefont{Lian, Wang, Sun, Vaezi,
  and Zhang}}]{lian2018quantum}
\bibinfo{author}{\bibfnamefont{B.}~\bibnamefont{Lian}},
  \bibinfo{author}{\bibfnamefont{J.}~\bibnamefont{Wang}},
  \bibinfo{author}{\bibfnamefont{X.-Q.} \bibnamefont{Sun}},
  \bibinfo{author}{\bibfnamefont{A.}~\bibnamefont{Vaezi}}, \bibnamefont{and}
  \bibinfo{author}{\bibfnamefont{S.-C.} \bibnamefont{Zhang}},
  \bibinfo{journal}{Physical Review B} \textbf{\bibinfo{volume}{97}},
  \bibinfo{pages}{125408} (\bibinfo{year}{2018}).

\bibitem[{\citenamefont{Ji and Wen}(2018)}]{ji20181}
\bibinfo{author}{\bibfnamefont{W.}~\bibnamefont{Ji}} \bibnamefont{and}
  \bibinfo{author}{\bibfnamefont{X.-G.} \bibnamefont{Wen}},
  \bibinfo{journal}{Physical review letters} \textbf{\bibinfo{volume}{120}},
  \bibinfo{pages}{107002} (\bibinfo{year}{2018}).

\bibitem[{\citenamefont{Huang et~al.}(2018)\citenamefont{Huang, Setiawan, and
  Sau}}]{huang2018disorder}
\bibinfo{author}{\bibfnamefont{Y.}~\bibnamefont{Huang}},
  \bibinfo{author}{\bibfnamefont{F.}~\bibnamefont{Setiawan}}, \bibnamefont{and}
  \bibinfo{author}{\bibfnamefont{J.~D.} \bibnamefont{Sau}},
  \bibinfo{journal}{Physical Review B} \textbf{\bibinfo{volume}{97}},
  \bibinfo{pages}{100501} (\bibinfo{year}{2018}).

\bibitem[{\citenamefont{Lian et~al.}(2017)\citenamefont{Lian, Sun, Vaezi, Qi,
  and Zhang}}]{lian2017non}
\bibinfo{author}{\bibfnamefont{B.}~\bibnamefont{Lian}},
  \bibinfo{author}{\bibfnamefont{X.-Q.} \bibnamefont{Sun}},
  \bibinfo{author}{\bibfnamefont{A.}~\bibnamefont{Vaezi}},
  \bibinfo{author}{\bibfnamefont{X.-L.} \bibnamefont{Qi}}, \bibnamefont{and}
  \bibinfo{author}{\bibfnamefont{S.-C.} \bibnamefont{Zhang}},
  \bibinfo{journal}{arXiv preprint arXiv:1712.06156}  (\bibinfo{year}{2017}).

\bibitem[{\citenamefont{Wang et~al.}(2013)\citenamefont{Wang, Lian, Zhang, Xu,
  and Zhang}}]{wang2013quantum}
\bibinfo{author}{\bibfnamefont{J.}~\bibnamefont{Wang}},
  \bibinfo{author}{\bibfnamefont{B.}~\bibnamefont{Lian}},
  \bibinfo{author}{\bibfnamefont{H.}~\bibnamefont{Zhang}},
  \bibinfo{author}{\bibfnamefont{Y.}~\bibnamefont{Xu}}, \bibnamefont{and}
  \bibinfo{author}{\bibfnamefont{S.-C.} \bibnamefont{Zhang}},
  \bibinfo{journal}{Physical review letters} \textbf{\bibinfo{volume}{111}},
  \bibinfo{pages}{136801} (\bibinfo{year}{2013}).

\bibitem[{\citenamefont{Fang et~al.}(2014)\citenamefont{Fang, Gilbert, and
  Bernevig}}]{fang2014large}
\bibinfo{author}{\bibfnamefont{C.}~\bibnamefont{Fang}},
  \bibinfo{author}{\bibfnamefont{M.~J.} \bibnamefont{Gilbert}},
  \bibnamefont{and} \bibinfo{author}{\bibfnamefont{B.~A.}
  \bibnamefont{Bernevig}}, \bibinfo{journal}{Phys. Rev. Lett.}
  \textbf{\bibinfo{volume}{112}}, \bibinfo{pages}{046801}
  (\bibinfo{year}{2014}),
  \urlprefix\url{https://link.aps.org/doi/10.1103/PhysRevLett.112.046801}.

\bibitem[{\citenamefont{Wang and Lian}(2018)}]{wang2018multiple}
\bibinfo{author}{\bibfnamefont{J.}~\bibnamefont{Wang}} \bibnamefont{and}
  \bibinfo{author}{\bibfnamefont{B.}~\bibnamefont{Lian}},
  \bibinfo{journal}{arXiv preprint arXiv:1805.10763}  (\bibinfo{year}{2018}).

\bibitem[{\citenamefont{Chang et~al.}(2013)\citenamefont{Chang, Zhang, Feng,
  Shen, Zhang, Guo, Li, Ou, Wei, Wang et~al.}}]{chang2013experimental}
\bibinfo{author}{\bibfnamefont{C.-Z.} \bibnamefont{Chang}},
  \bibinfo{author}{\bibfnamefont{J.}~\bibnamefont{Zhang}},
  \bibinfo{author}{\bibfnamefont{X.}~\bibnamefont{Feng}},
  \bibinfo{author}{\bibfnamefont{J.}~\bibnamefont{Shen}},
  \bibinfo{author}{\bibfnamefont{Z.}~\bibnamefont{Zhang}},
  \bibinfo{author}{\bibfnamefont{M.}~\bibnamefont{Guo}},
  \bibinfo{author}{\bibfnamefont{K.}~\bibnamefont{Li}},
  \bibinfo{author}{\bibfnamefont{Y.}~\bibnamefont{Ou}},
  \bibinfo{author}{\bibfnamefont{P.}~\bibnamefont{Wei}},
  \bibinfo{author}{\bibfnamefont{L.-L.} \bibnamefont{Wang}},
  \bibnamefont{et~al.}, \bibinfo{journal}{Science} p. \bibinfo{pages}{1232003}
  (\bibinfo{year}{2013}).

\bibitem[{Jeo()}]{Jeon-SM}
\bibinfo{note}{See Supplemental Material which includes the discussion for
  matrix element of BdG hamiltonian, semimetallic regions, continuum limit, and
  phase diagrams for periodic $h/2e$ or $h/e$ fluxoids of a hexagonal shape in
  triangular lattice.}

\bibitem[{\citenamefont{Thouless et~al.}(1982)\citenamefont{Thouless, Kohmoto,
  Nightingale, and den Nijs}}]{Thouless82}
\bibinfo{author}{\bibfnamefont{D.~J.} \bibnamefont{Thouless}},
  \bibinfo{author}{\bibfnamefont{M.}~\bibnamefont{Kohmoto}},
  \bibinfo{author}{\bibfnamefont{M.~P.} \bibnamefont{Nightingale}},
  \bibnamefont{and} \bibinfo{author}{\bibfnamefont{M.}~\bibnamefont{den Nijs}},
  \bibinfo{journal}{Phys. Rev. Lett.} \textbf{\bibinfo{volume}{49}},
  \bibinfo{pages}{405} (\bibinfo{year}{1982}),
  \urlprefix\url{http://link.aps.org/doi/10.1103/PhysRevLett.49.405}.

\bibitem[{\citenamefont{Mong et~al.}(2014)\citenamefont{Mong, Clarke, Alicea,
  Lindner, Fendley, Nayak, Oreg, Stern, Berg, Shtengel et~al.}}]{Mong14}
\bibinfo{author}{\bibfnamefont{R.~S.~K.} \bibnamefont{Mong}},
  \bibinfo{author}{\bibfnamefont{D.~J.} \bibnamefont{Clarke}},
  \bibinfo{author}{\bibfnamefont{J.}~\bibnamefont{Alicea}},
  \bibinfo{author}{\bibfnamefont{N.~H.} \bibnamefont{Lindner}},
  \bibinfo{author}{\bibfnamefont{P.}~\bibnamefont{Fendley}},
  \bibinfo{author}{\bibfnamefont{C.}~\bibnamefont{Nayak}},
  \bibinfo{author}{\bibfnamefont{Y.}~\bibnamefont{Oreg}},
  \bibinfo{author}{\bibfnamefont{A.}~\bibnamefont{Stern}},
  \bibinfo{author}{\bibfnamefont{E.}~\bibnamefont{Berg}},
  \bibinfo{author}{\bibfnamefont{K.}~\bibnamefont{Shtengel}},
  \bibnamefont{et~al.}, \bibinfo{journal}{Phys. Rev. X}
  \textbf{\bibinfo{volume}{4}}, \bibinfo{pages}{011036} (\bibinfo{year}{2014}),
  \urlprefix\url{https://link.aps.org/doi/10.1103/PhysRevX.4.011036}.

\bibitem[{\citenamefont{Rickhaus et~al.}(2012)\citenamefont{Rickhaus, Weiss,
  Marot, and Schönenberger}}]{Rickhaus12}
\bibinfo{author}{\bibfnamefont{P.}~\bibnamefont{Rickhaus}},
  \bibinfo{author}{\bibfnamefont{M.}~\bibnamefont{Weiss}},
  \bibinfo{author}{\bibfnamefont{L.}~\bibnamefont{Marot}}, \bibnamefont{and}
  \bibinfo{author}{\bibfnamefont{C.}~\bibnamefont{Schönenberger}},
  \bibinfo{journal}{Nano Letters} \textbf{\bibinfo{volume}{12}},
  \bibinfo{pages}{1942} (\bibinfo{year}{2012}), \bibinfo{note}{pMID: 22417183},
  \eprint{http://dx.doi.org/10.1021/nl204415s},
  \urlprefix\url{http://dx.doi.org/10.1021/nl204415s}.

\bibitem[{\citenamefont{Amet et~al.}(2016)\citenamefont{Amet, Ke, Borzenets,
  Wang, Watanabe, Taniguchi, Deacon, Yamamoto, Bomze, Tarucha et~al.}}]{Amet16}
\bibinfo{author}{\bibfnamefont{F.}~\bibnamefont{Amet}},
  \bibinfo{author}{\bibfnamefont{C.~T.} \bibnamefont{Ke}},
  \bibinfo{author}{\bibfnamefont{I.~V.} \bibnamefont{Borzenets}},
  \bibinfo{author}{\bibfnamefont{J.}~\bibnamefont{Wang}},
  \bibinfo{author}{\bibfnamefont{K.}~\bibnamefont{Watanabe}},
  \bibinfo{author}{\bibfnamefont{T.}~\bibnamefont{Taniguchi}},
  \bibinfo{author}{\bibfnamefont{R.~S.} \bibnamefont{Deacon}},
  \bibinfo{author}{\bibfnamefont{M.}~\bibnamefont{Yamamoto}},
  \bibinfo{author}{\bibfnamefont{Y.}~\bibnamefont{Bomze}},
  \bibinfo{author}{\bibfnamefont{S.}~\bibnamefont{Tarucha}},
  \bibnamefont{et~al.}, \bibinfo{journal}{Science}
  \textbf{\bibinfo{volume}{352}}, \bibinfo{pages}{966} (\bibinfo{year}{2016}),
  ISSN \bibinfo{issn}{0036-8075},
  \eprint{http://science.sciencemag.org/content/352/6288/966.full.pdf},
  \urlprefix\url{http://science.sciencemag.org/content/352/6288/966}.

\bibitem[{\citenamefont{{Ben Shalom} et~al.}(2016)\citenamefont{{Ben Shalom},
  Zhu, Fal'ko, Mishchenko, Kretinin, Novoselov, Woods, Watanabe, Taniguchi,
  Geim et~al.}}]{Shalom16}
\bibinfo{author}{\bibfnamefont{M.}~\bibnamefont{{Ben Shalom}}},
  \bibinfo{author}{\bibfnamefont{M.~J.} \bibnamefont{Zhu}},
  \bibinfo{author}{\bibfnamefont{V.~I.} \bibnamefont{Fal'ko}},
  \bibinfo{author}{\bibfnamefont{A.}~\bibnamefont{Mishchenko}},
  \bibinfo{author}{\bibfnamefont{A.~V.} \bibnamefont{Kretinin}},
  \bibinfo{author}{\bibfnamefont{K.~S.} \bibnamefont{Novoselov}},
  \bibinfo{author}{\bibfnamefont{C.~R.} \bibnamefont{Woods}},
  \bibinfo{author}{\bibfnamefont{K.}~\bibnamefont{Watanabe}},
  \bibinfo{author}{\bibfnamefont{T.}~\bibnamefont{Taniguchi}},
  \bibinfo{author}{\bibfnamefont{A.~K.} \bibnamefont{Geim}},
  \bibnamefont{et~al.}, \bibinfo{journal}{Nat. Phys.}
  \textbf{\bibinfo{volume}{12}}, \bibinfo{pages}{318} (\bibinfo{year}{2016}),
  ISSN \bibinfo{issn}{1745-2473},
  \urlprefix\url{http://dx.doi.org/10.1038/nphys3592 http://10.0.4.14/nphys3592
  https://www.nature.com/articles/nphys3592#supplementary-information
  http://www.nature.com/articles/nphys3592}.

\bibitem[{\citenamefont{Lee et~al.}({2017})\citenamefont{Lee, Huang, Efetov,
  Wei, Hart, Taniguchi, Watanabe, Yacoby, and Kim}}]{Lee17}
\bibinfo{author}{\bibfnamefont{G.-H.} \bibnamefont{Lee}},
  \bibinfo{author}{\bibfnamefont{K.-F.} \bibnamefont{Huang}},
  \bibinfo{author}{\bibfnamefont{D.~K.} \bibnamefont{Efetov}},
  \bibinfo{author}{\bibfnamefont{D.~S.} \bibnamefont{Wei}},
  \bibinfo{author}{\bibfnamefont{S.}~\bibnamefont{Hart}},
  \bibinfo{author}{\bibfnamefont{T.}~\bibnamefont{Taniguchi}},
  \bibinfo{author}{\bibfnamefont{K.}~\bibnamefont{Watanabe}},
  \bibinfo{author}{\bibfnamefont{A.}~\bibnamefont{Yacoby}}, \bibnamefont{and}
  \bibinfo{author}{\bibfnamefont{P.}~\bibnamefont{Kim}},
  \bibinfo{journal}{{Nat. Phys.}} \textbf{\bibinfo{volume}{{13}}},
  \bibinfo{pages}{{693}} (\bibinfo{year}{{2017}}), ISSN
  \bibinfo{issn}{{1745-2473}}.

\bibitem[{Sah()}]{Sahu18}
\bibinfo{note}{M. R. Sahu, X. Liu, A. K. Paul, S. Das, P. Raychaudhuri, J. K.
  Jain, and A. Das, unpublished.}

\bibitem[{\citenamefont{Buzdin}(2005)}]{buzdin2005proximity}
\bibinfo{author}{\bibfnamefont{A.~I.} \bibnamefont{Buzdin}},
  \bibinfo{journal}{Reviews of modern physics} \textbf{\bibinfo{volume}{77}},
  \bibinfo{pages}{935} (\bibinfo{year}{2005}).

\bibitem[{\citenamefont{Bergeret et~al.}(2005)\citenamefont{Bergeret, Volkov,
  and Efetov}}]{bergeret2005odd}
\bibinfo{author}{\bibfnamefont{F.}~\bibnamefont{Bergeret}},
  \bibinfo{author}{\bibfnamefont{A.~F.} \bibnamefont{Volkov}},
  \bibnamefont{and} \bibinfo{author}{\bibfnamefont{K.~B.}
  \bibnamefont{Efetov}}, \bibinfo{journal}{Reviews of modern physics}
  \textbf{\bibinfo{volume}{77}}, \bibinfo{pages}{1321} (\bibinfo{year}{2005}).

\bibitem[{\citenamefont{Eschrig}(2011)}]{eschrig2011spin}
\bibinfo{author}{\bibfnamefont{M.}~\bibnamefont{Eschrig}},
  \bibinfo{journal}{Phys. Today} \textbf{\bibinfo{volume}{64}},
  \bibinfo{pages}{43} (\bibinfo{year}{2011}).

\bibitem[{\citenamefont{Eschrig}(2015)}]{eschrig2015spin}
\bibinfo{author}{\bibfnamefont{M.}~\bibnamefont{Eschrig}},
  \bibinfo{journal}{Reports on Progress in Physics}
  \textbf{\bibinfo{volume}{78}}, \bibinfo{pages}{104501}
  (\bibinfo{year}{2015}).

\bibitem[{\citenamefont{Lian et~al.}(2016)\citenamefont{Lian, Wang, and
  Zhang}}]{lian2016edge}
\bibinfo{author}{\bibfnamefont{B.}~\bibnamefont{Lian}},
  \bibinfo{author}{\bibfnamefont{J.}~\bibnamefont{Wang}}, \bibnamefont{and}
  \bibinfo{author}{\bibfnamefont{S.-C.} \bibnamefont{Zhang}},
  \bibinfo{journal}{Physical Review B} \textbf{\bibinfo{volume}{93}},
  \bibinfo{pages}{161401} (\bibinfo{year}{2016}).

\bibitem[{\citenamefont{Bernevig and Hughes}(2013)}]{Bernevig13}
\bibinfo{author}{\bibfnamefont{B.}~\bibnamefont{Bernevig}} \bibnamefont{and}
  \bibinfo{author}{\bibfnamefont{T.}~\bibnamefont{Hughes}},
	  \bibinfo{title}{
  \emph{
		  Topological Insulators and Topological Superconductors
  } 
	  }
  (\bibinfo{publisher}{Princeton University Press},
  \bibinfo{year}{2013}).

\end{thebibliography}
\bibliographystyle{apsrev}
\end{document}